\documentclass[a4paper]{article}
\usepackage{latexsym}
\usepackage{amsbsy}
\usepackage{amssymb}
\usepackage{amsmath}
\usepackage{graphicx}
\usepackage{epsfig}
\usepackage[latin1]{inputenc}
\usepackage{geometry}
\usepackage{caption}
\usepackage{epstopdf}

\input{amssym.def}
\input{amssym}
\newtheorem{thm}{Theorem}[section]
\newtheorem{defin}[thm]{Definition}
\newtheorem{lem}[thm]{Lemma}
\newtheorem{prop}[thm]{Proposition}
\newtheorem{cor}[thm]{Corollary}
\newtheorem{rem}[thm]{Remark}
\newtheorem{ex}[thm]{Example}
\newcommand{\bthm}{\begin{thm}}
\newcommand{\ethm}{\end{thm}}
\newcommand{\bd}{\begin{defin}}
\newcommand{\ed}{\end{defin}}
\newcommand{\blem}{\begin{lem}}
\newcommand{\elem}{\end{lem}}
\newcommand{\bcor}{\begin{cor}}
\newcommand{\ecor}{\end{cor}}
\newcommand{\bprop}{\begin{prop}}
\newcommand{\eprop}{\end{prop}}
\newcommand{\brem}{\begin{rem} \rm}
\newcommand{\erem}{\end{rem}}
\newcommand{\bex}{\begin{ex} \rm}
\newcommand{\eex}{\end{ex}}

\newcommand{\beq}{\begin{equation}}
\newcommand{\eeq}{\end{equation}}
\newcommand{\bea}{\begin{eqnarray}}
\newcommand{\eea}{\end{eqnarray}}
\newcommand{\beas}{\begin{eqnarray*}}
\newcommand{\eeas}{\end{eqnarray*}}
\newcommand{\beqs}{\begin{equation*}}
\newcommand{\eeqs}{\end{equation*}}
\newcommand{\bi}{\begin{itemize}}
\newcommand{\ei}{\end{itemize}}
\newcommand{\ben}{\begin{enumerate}}
\newcommand{\een}{\end{enumerate}}
\newcommand{\ba}{\begin{array}}
\newcommand{\ea}{\end{array}}

\newcommand{\vphi}{\varphi}

\newcommand{\tg}{\mathop{\rm tg}\nolimits}

\newcommand{\notmid}{\mid\kern-0.5em\not\kern0.5em}

\def\dj{d\kern-0.4em\char"16\kern-0.1em}
\def\Dj{\mbox{\raise0.3ex\hbox{-}\kern-0.4em D}}
\input{tcilatex}
\begin{document}

\title{Distributed order fractional constitutive stress-strain relation in
wave propagation modeling}
\author{Sanja Konjik\thanks{
Department of Mathematics and Informatics, Faculty of Sciences, University
of Novi Sad, Trg Dositeja Obradovi\'ca 4, 21000 Novi Sad, Serbia, Electronic
mail: sanja.konjik@dmi.uns.ac.rs}, Ljubica Oparnica\thanks{
Faculty of Education, University of Novi Sad, Podgori\v cka 4, 25000 Sombor,
Serbia, Electronic mail: ljubica.oparnica@pef.uns.ac.rs}, Du\v san Zorica%
\thanks{
Mathematical Institute, Serbian Academy of Arts and Sciences, Kneza Mihaila
36, 11000 Belgrade, Serbia, and Department of Physics, Faculty of Sciences,
University of Novi Sad, Trg Dositeja Obradovi\'ca 4, 21000 Novi Sad, Serbia,
Electronic mail: dusan\textunderscore zorica@mi.sanu.ac.rs} }
\date{}
\maketitle

\begin{abstract}
\noindent Distributed order fractional model of viscoelastic body is used in
order to describe wave propagation in infinite media. Existence and
uniqueness of fundamental solution to the generalized Cauchy problem,
corresponding to fractional wave equation, is studied. The explicit form of
fundamental solution is calculated, and wave propagation speed, arising from
solution's support, is found to be connected with the material properties at
initial time instant. Existence and uniqueness of the fundamental solutions
to the fractional wave equations corresponding to four thermodynamically
acceptable classes of linear fractional constitutive models, as well as to
power type distributed order model, are established and explicit forms of
the corresponding fundamental solutions are obtained. 

\end{abstract}



\section{Introduction}

\label{sec:intro} 

The aim is to solve and analyze the wave equation for infinite
one-dimensional viscoelastic media, described by the distributed order
fractional constitutive model. More precisely, we shall study the following
system of equations: 
\begin{eqnarray}
\frac{\partial }{\partial x}\sigma (x,t) &=&\rho \,\frac{\partial ^{2}}{%
\partial t^{2}}u(x,t),  \label{eq-mot} \\
\int_{0}^{1}\phi _{\sigma }(\alpha )\,{}_{0}D_{t}^{\alpha }\sigma
(x,t)\,d\alpha &=&E\,\int_{0}^{1}\phi _{\varepsilon }(\alpha
)\,{}_{0}D_{t}^{\alpha }\varepsilon (x,t)\,d\alpha ,  \label{const-eq} \\
\varepsilon (x,t) &=&\frac{\partial }{\partial x}u(x,t),  \label{strain}
\end{eqnarray}%
where $u$, $\sigma $, and $\varepsilon $ are displacement, stress, and
strain, $x\in \mathbb{R}$, $t>0$, $\rho =\mbox{const.}$ is the material
density, $E=\mbox{const.}$ is the generalized Young modulus of elasticity,
while $\phi _{\sigma }$ and $\phi _{\varepsilon }$ are constitutive
functions or distributions, describing material properties. The left (right)
hand side of the second equation is a distributed order fractional
derivative of $\sigma $ ($\varepsilon $), with ${}_{0}D_{t}^{\alpha }$ being
Riemann-Liouville fractional derivative of order $\alpha \in \lbrack 0,1),$
to be defined in Section \ref{sec:set&tol}.

The first equation is the equation of motion of one-dimensional deformable
body, the second equation is the constitutive equation of distributed order
fractional type, describing the mechanical properties of the linear
viscoelastic body, and the third equation is the strain for small local
deformations. Originating from the basic equations of the continuum
mechanics, see \cite{a-g}, the equation of motion and strain are preserved,
since they hold true for any type of deformable body and only the
constitutive equation, which is the Hooke law for an elastic body, is
changed by distributed order fractional model, thus adapted for viscoelastic
type media.

The distributed order constitutive stress-strain relation (\ref{const-eq})
generalizes integer and fractional order constitutive models of linear
viscoelasticity 
\begin{equation}
\sum_{i=1}^{n}a_{i}\,{}_{0}D_{t}^{\alpha _{i}}\sigma
(x,t)=E\,\sum_{j=1}^{m}b_{i}\,{}_{0}D_{t}^{\beta _{j}}\varepsilon (x,t),
\label{LinFractCE}
\end{equation}%
having differentiation orders up to the first order, if the constitutive
distributions $\phi _{\sigma }$ and $\phi _{\varepsilon }$ are chosen as 
\begin{equation*}
\phi _{\sigma }(\alpha ):=\sum_{i=1}^{n}a_{i}\delta (\alpha -\alpha
_{i}),\quad \phi _{\varepsilon }(\alpha ):=\sum_{j=1}^{m}b_{i}\delta (\alpha
-\beta _{j}),
\end{equation*}%
with model parameters: $a_{i},b_{j}>0$ as generalized time constants, and $%
\alpha _{i},\beta _{j}\in \lbrack 0,1)$ as orders of fractional
differentiation, where $i=1,\ldots ,n$, $j=1,\ldots ,m$. Thermodynamical
consistency of linear fractional constitutive equation \eqref{LinFractCE} is
examined in \cite{AKOZ}, where it is shown that there are four cases of %
\eqref{LinFractCE} when the restrictions on model parameters guarantee its
thermodynamical consistency. The power type distributed order model of
viscoelastic body 
\begin{equation}
\int_{0}^{1}a^{\alpha }\,{}_{0}D_{t}^{\alpha }\sigma (x,t)\,d\alpha
=E\,\int_{0}^{1}b^{\alpha }\,{}_{0}D_{t}^{\alpha }\varepsilon (x,t)\,d\alpha
,  \label{DOCE}
\end{equation}%
considered in \cite{a-2003} and revisited in \cite{AKOZ}, is obtained if the
constitutive functions $\phi _{\sigma }$ and $\phi _{\varepsilon }$ are
chosen as%
\begin{equation*}
\phi _{\sigma }(\alpha ):=a^{\alpha },\quad \phi _{\varepsilon }(\alpha
):=b^{\alpha },
\end{equation*}%
with the time constants $a,b>0$ satisfying $a\leq b$ guaranteeing model's
dimensional homogeneity and thermodynamical consistency.

The aim is to consider the wave equation, written as system of equations (%
\ref{eq-mot}) - (\ref{strain}), on infinite domain for general form of the
distributed order constitutive equation (\ref{const-eq}), i.e., to establish
the existence of its solution, to calculate its fundamental solution, and to
discuss the wave propagation speed using solution's support properties. In
order to prove the existence of fundamental solution, three assumptions are
needed, with two of them guaranteeing the solvability of the constitutive
equation with respect to both stress and strain, while the third one ensures
fundamental solution's decay at infinity. In order to calculate the
fundamental solution, another three assumptions are needed, guaranteeing the
validity of derived formula. One of the assumptions is closely related to
the support property of fundamental solution, implying the connection of
wave propagation speed with the material properties. More precisely, the
wave propagation speed is proportional to the square root of the glass
modulus. The other assumption restricts the growth of creep compliance for
large times.

Although assumptions might not seem straightforward to be verified, all four
classes of thermodynamically consistent models, corresponding to the linear
fractional constitutive equation \eqref{LinFractCE}, are proved to satisfy
all six assumptions, implying the existence and explicit form of the
fundamental solution to all four classes of thermodynamically admissible
models of linear viscoelasticity, having the orders of differentiation in
interval $[0,1).$ The power type distributed order constitutive model %
\eqref{DOCE}, as an genuine distributed order model, is proved to comply
with all six assumptions as well. Numerical examples are chosen so that each
class of linear fractional models, except for the one already analyzed in 
\cite{KOZ10,KOZ11}, as well as the power type distributed order model, are
represented.

Generalizations of the classical wave equation are considered on infinite
domain in \cite{KOZ10,KOZ11}, where the constitutive equation, representing
a single class of thermodynamically consistent linear fractional
constitutive equation \eqref{LinFractCE}, is chosen to be either the
fractional Zener model, or its generalization having arbitrary number of
fractional derivatives of same orders acting on both stress and strain. Wave
propagation speed is obtained from the support property of fundamental
solution in \cite{KOZ10,KOZ11}, while in \cite{HOZ16} tools of microlocal
analysis were employed in order to examine the singularity propagation
properties in the case of fractional Zener wave equation. Fractional wave
equation, with power type distributed order model \eqref{DOCE} as the
constitutive equation, is considered on finite domain in \cite{APZ-4,APZ-3}.
The fractional Zener wave equation on the bounded domain is analyzed in \cite%
{R-S}, while modified Zener and modified Maxwell wave equations are
considered for bounded and semi-bounded domains in \cite{R-S1,R-S2}. Several
other problems involving generalizations of the classical wave equations are
reviewed in \cite{APSZ-2,R-S-2010}.

The problem of wave propagation in viscoelastic solids of integer and
fractional order type, described by the linear fractional model (\ref%
{LinFractCE}), is considered in \cite{CaputoMainardi-1971a} and the
fractional Zener model is introduced and analyzed and in \cite%
{CaputoMainardi-1971b}, where the wave speed is obtained for the linear
viscoelastic solids of integer order and where dissipation properties of the
Zener fractional wave equation are analyzed. The Buchen-Mainardi wavefront
expansion of solution to the wave equation for viscoelastic body near the
wavefront is introduced in \cite{BuchenMainardi} and used in \cite%
{ColombaroGiustiMainardi1}, where integer and fractional order Maxwell and
Kelvin-Voigt models of viscoelastic materials are considered, while
viscoelastic Bessel medium is considered in \cite{ColombaroGiustiMainardi2}.
The Bessel constitutive equation is introduced in \cite{GiustiMainardi} and
further analyzed in \cite{ColombaroGiustiMainardi}. Asymptotic behavior of
viscoelastic wave equation fundamental's solution near the wavefront is
examined in \cite{Han6}, while dispersion, attenuation and wavefronts
corresponding to specific viscoelastic materials are examined in \cite%
{Han7,Han8}. The Achenbach and Reddy method of discontinuities is used in 
\cite{Hazanov} in order to study the shock wave propagation in materials
with memory. An extensive review of the viscoelasticity of fractional order
and of dispersion and attenuation properties of corresponding wave
equations, including the wavefront expansions, can be found in \cite{Mai-10}.

Fractional diffusion-wave equation is known to generalize both diffusion
(heat conduction) equation and wave equation, due to the fractional
derivative of order from interval $\left( 1,2\right) $ instead of the first
and second order derivatives appearing in the classical equations. This
equation is introduced and solved for the Cauchy and signalling problems in 
\cite{Mai96,Mai96-1}, reinterpreted in \cite{MainardiParadisi} within the
theory of linear viscoelasticity of fractional order, while the propagation
speed of the maximum of fundamental solution is examined in \cite%
{LuchkoMainardi-1,LuchkoMainardi-2,LuchkoMainardiPovstenko}.

Considerations regarding the different forms of wave equations for the
viscoelastic media can be found in \cite%
{CaputoCarcione,EndeLionLammering,Gaul,GeorgievRubinoSampalmieri,Han0,Han3,Han9,Han2,HolmHolm,MustafaMessaoudi,NH}%
, including derivation and solution of a model describing propagation and
attenuation of two-dimensional dilatational waves; derivation of
thermodynamically consistent fractional wave equation which allows for the
calculation of LAMB waves; analysis of vibrations damping; proving the
existence of global solution to memory-type wave equation; application of
Volterra's and Lokshin's theory in studying well-posedness and regularity of
solutions to wave equations; examining step signal propagation in hereditary
medium; formulating and analyzing the fundamental solution to the
anisotropic multi-dimensional linear viscoelastic wave equation; formulating
Hamiltonian and Lagrangian theory of viscoelasticity; formulating
restrictions that allow the complete monotonicity; general stability result;
and analysis of the Zener fractional wave equation, respectively. Wave
equation obtained in modelling string vibration is analyzed in \cite%
{SandevTomovski,TomovskiSandev}.

Wave equation for viscoelastic media has extensive applicability in seismic
pulse propagation modeling, see \cite%
{JMCarcione,CarcioneCavalliniMainardiHanyga,Mai97,ZhuCarcione}. The review
of seismic wave propagation in real media is given in \cite{Carcione}. The
modeling of acoustic pulse propagation in viscoelastic media, including
biological materials, also involves the use of fractional generalizations of
the wave equation, usually used in the frequency domain, as done in \cite%
{CaputoCarcioneCavallini,ChenHolm,FellahDepollier,FellahDepollierFellah-1,FellahDepollierFellah,HolmNasholm,HolmNasholmPrieurSinkus,HolmSinkus,NasholmHolm,SebaaFellahLauriksDepollier}%
.

The exposition of article is organized as follows. In Section \ref%
{sec:set&tol}, we introduce and explain the setting and tools that will be
used in our study. The theoretical part of our work is presented in Section %
\ref{sec:dofzwe}, where we derive conditions that: yield equivalence of
system \eqref{system-1} - \eqref{system-3} with the wave equation for
distributed order type viscoelastic media \eqref{dowe}; provide existence
and uniqueness of a fundamental solution to the generalized Cauchy problem
associated to the wave equation \eqref{dowe}; and guarantee explicit
calculation of the solution. The general theory, developed in Section \ref%
{sec:dofzwe}, is further analyzed and applied to all four classes of
thermodynamically admissible linear fractional models of viscoelastic body
in Section \ref{sec:lfmS}, as well as to the power type distributed order
model in Section \ref{sec:cm}. As a verification of obtained theoretical
results, in Section \ref{sec:nums}, we examine illustrative numerical
examples, using modified Zener, modified Maxwell, and Maxwell models as the
representatives of thermodynamically admissible classes of linear fractional
models, as well as the power type distributed order model.


\section{Setting and tools}

\label{sec:set&tol} 

Distributed order fractional model of the wave equation in viscoelastic
infinite media will be studied in a distributional framework. We shall be
dealing with the spaces of distributions $\mathcal{D} ^{\prime }(\Omega)$, $%
\Omega \subseteq \mathbb{R}$ is open, tempered distributions $\mathcal{S}
^{\prime }(\mathbb{R})$, and their subspaces $\mathcal{D} ^{\prime }_{+}(%
\mathbb{R})$ and $\mathcal{S} ^{\prime }_{+}(\mathbb{R})$ consisting of
distributions supported in $[0,\infty)$. Also, space $\mathcal{S} ^{\prime }(%
\mathbb{R}\times \mathbb{R}_{+})$ of all distributions $u\in \mathcal{S}
^{\prime }(\mathbb{R}^{2})$ vanishing for $t<0$, will be a part of the
framework. Similarly, by $L^1_{loc+} (\mathbb{R})$ we shall denote the space
of $L^1_{loc} (\mathbb{R})$-functions that vanish for $t<0$.

Taking $t\in [0,T]$, $T>0$, and $y\in AC([0,T])$, one defines the left
Riemann-Liouville fractional derivative of order $\alpha\in [0,1)$ as 
\begin{equation*}
{}_{0}D_{t}^{\alpha}y(t) =\frac{1}{\Gamma (1-\alpha)} \frac{d}{dt}
\int_{0}^{t} \frac{y(\tau)}{(t-\tau)^{\alpha}}\,d\tau,
\end{equation*}
where $\Gamma$ is the Euler gamma function. In the distributional setting,
the Riemann-Liouville fractional derivative can be seen as a convolution
operator $f_{-\alpha}\ast$, $\alpha\in[0,1)$, where the family $%
\{f_\alpha\}_{\alpha\in\mathbb{R}}\in\mathcal{S} ^{\prime }_+(\mathbb{R})$
is defined by 
\begin{equation*}
f_{\alpha}(t) = \left\{ 
\begin{array}{ll}
H(t) \frac{t^{\alpha -1}}{\Gamma (\alpha)}, & \alpha >0, \\ 
\frac{d^{N}}{dt^{N}}f_{\alpha +N}(t), & \alpha \leq 0, \alpha +N>0, N\in 
\mathbb{N},%
\end{array}
\right.
\end{equation*}
$H$ is the Heaviside function, $N$-th derivative is to be understood in the
distributional sense, and $f_{\alpha}:\mathcal{S} ^{\prime }_+\to\mathcal{S}
^{\prime }_+$. Moreover, for $y\in AC([0,T])$ we have that $%
{}_{0}D_{t}^{\alpha}y(t)$ and $f_\alpha\ast y$, $\alpha\in[0,1)$, coincide,
and the mappings $\alpha\mapsto f_\alpha\ast y:\mathbb{R }\to \mathcal{S}
^{\prime }(\mathbb{R})$ and $\alpha\mapsto \langle f_\alpha\ast y, \varphi
\rangle:\mathbb{R }\to \mathbb{R}$, $\varphi\in\mathcal{S} (\mathbb{R})$,
are smooth, cf. \cite{AOP,APSZ-2}.

For $\phi\in\mathcal{E} ^{\prime }(\mathbb{R})$ and $y\in\mathcal{S}
^{\prime }_+(\mathbb{R})$ one can define the distributed order fractional
derivative of $y$ as an element of $\mathcal{S} ^{\prime }_+(\mathbb{R})$ by 
\begin{equation}  \label{DefInt}
\Big\langle \int_{\mathop{\rm supp}\nolimits \phi}
\phi(\alpha){}_0D_t^{\alpha}y(t) \,d\alpha, \varphi(t) \Big\rangle: = %
\big\langle \phi(\alpha) , \langle {}_0D_t^{\alpha}y(t) , \varphi(t) \rangle %
\big\rangle, \quad \varphi\in\mathcal{S} (\mathbb{R}).
\end{equation}
When $\mathop{\rm supp}\nolimits\phi\subset [c,d]$ we shall write $\int_c^d
\phi(\alpha){}_0D_t^{\alpha}y(t) \,d\alpha$. Special cases are continuous
functions $\phi$ of $\alpha$ in $[c,d]$, or sums of the Dirac delta
distributions $\phi(\cdot)=\sum_{i=0}^{k} c_{i}\delta(\cdot -\alpha_{i})$, $%
\alpha_{i}\in \mathbb{R}$, $i\in\{0,1,...,k\}$, when the distributed order
fractional derivative reduces to a sum of finite numbers of
Riemann-Liouville fractional derivatives, i.e., $\sum_{i=0}^{k}c_{i}\cdot{}%
_0D_t^{\alpha_i}y(t)$, cf. \cite{AOP,APSZ-2}.

Basic tools for solving our problem will be the Fourier and Laplace integral
transforms. The Fourier transform is a continuous isomorphism on $\mathcal{S}
(\mathbb{R})$: For $\varphi \in \mathcal{S} (\mathbb{R})$ it is defined as 
\begin{equation*}
\mathcal{F} \varphi (\xi)=\hat{\varphi}(\xi)=\int_{-\infty }^{\infty}
\varphi (x) e^{-i\xi x} \,dx, \quad \xi \in \mathbb{R}.
\end{equation*}
This definition can be extended to $y \in \mathcal{S} ^{\prime }(\mathbb{R})$
in the standard way: $\langle \mathcal{F} y,\varphi \rangle =\langle y,%
\mathcal{F} \varphi \rangle$, $\varphi \in \mathcal{S} (\mathbb{R})$.

The Laplace transform is well defined for $y\in \mathcal{D} ^{\prime }_{+}(%
\mathbb{R})$ with the property $e^{-\xi t}y\in\mathcal{S} ^{\prime }(\mathbb{%
R})$, for all $\xi >a$ (for $a=0$ it is the space $\mathcal{S} ^{\prime }_+(%
\mathbb{R})$) by 
\begin{equation}  \label{L-dd}
\mathcal{L} y(s) =\tilde{y}(s) = \mathcal{F} (e^{-\xi t}y)(\eta)=\langle y,
e^{-st}\rangle, \quad s=\xi +i\eta.
\end{equation}
It is holomorphic in the half plane $\mathop{\rm Re}\nolimits s>a$.
Expression $\langle y, e^{-st}\rangle$ has to be understood in the sense of
duality of $\mathcal{S} ^{\prime }(\mathbb{R})$ and $\mathcal{S} (\mathbb{R}%
) $: For a test function $\phi\in\mathcal{D} ^{\prime }_{+}(\mathbb{R})$
that is equal to $1$ in a neighborhood of the support of $y$, we have that $%
e^{-\xi t}y\in\mathcal{S} ^{\prime }(\mathbb{R})$ and $e^{-(s-\xi) t}\psi\in%
\mathcal{S} (\mathbb{R})$, for all $\xi >a>0$, thus $\langle y,
e^{-st}\rangle:=\langle e^{-\xi t}y, e^{-(s-\xi) t}\psi \rangle$. Also, for $%
y\in \mathcal{E} ^{\prime }(\mathbb{R})$ being a distribution with compact
support, \eqref{L-dd} is well-defined and holomorphic in $\mathbb{C}$.

If $y\in L^{1}_{loc}(\mathbb{R})$, $y(t)=0$ for $t<0$, and $|e^{-\xi
t}y(t)|\leq Pol(t)$, for all $\xi >a$, where $Pol$ denotes a polynomial in $%
t $, $t>0$, then 
\begin{equation*}
\mathcal{L} y(s)=\int_{0}^{\infty} y(t)e^{-st} \,dt, \quad \mathop{\rm Re}%
\nolimits s>a.
\end{equation*}

The inverse Laplace transform is well defined for holomorphic functions $Y$
in the half plane $\mathop{\rm Re}\nolimits s>a$ that satisfy 
\begin{equation}  \label{cond-inv-L}
|Y(s)| \leq A\frac{(1+|s|)^{m}}{|\mathop{\rm Re}\nolimits s|^k}, \qquad
m,k\in \mathbb{R},
\end{equation}
by 
\begin{equation*}
\mathcal{L} ^{-1}Y(t) =\frac{1}{2\pi i}\int_{a-i\infty}^{a+i\infty} Y(s)
e^{st}\, ds \in \mathcal{S} ^{\prime }_{+}(\mathbb{R}).
\end{equation*}


We list some properties of the Fourier and Laplace transforms that hold in
the distributional setting and will be used later: 
\begin{equation*}
\mathcal{F} (y^{(n)}(x)) (\xi) = (i\xi)^{n}\mathcal{F} y(\xi), \qquad 
\mathcal{L} [{}_{0}D_{t}^{\alpha}y](s) = s^{\alpha}\mathcal{L} y(s),
\end{equation*}
\begin{equation*}
\mathcal{L} [y_{1}\ast y_{2}(t)](s) = \mathcal{L} y_{1}(s)\cdot \mathcal{L}
y_{2}(s), \qquad \mathcal{L} \delta(s) = 1,
\end{equation*}
\begin{equation*}
{\mathcal{L} }\Big(\int_{\mathop{\rm supp}\nolimits \phi}
\phi(\alpha){}_0D_t^{\alpha}y(t)\,d\alpha \Big)(s) = \tilde{y}(s)\langle
\phi(\alpha) , s^{\alpha} \rangle, \qquad \mathop{\rm Re}\nolimits s>0,
\end{equation*}
and refer to \cite{DautryLions-vol5,vlad1,vlad2} for more details on the
Fourier and Laplace transforms.

Equation \eqref{dowe} together with initial conditions \eqref{ic} can be
considered as a Cauchy problem for the second order linear fractional
integro-differential operator with constant coefficients. In general, the
Cauchy problem for the classical wave operator $P$ is given by 
\begin{equation}  \label{mp-cp}
Pu(x,t)=f(x,t), \qquad u(x,0)=u_{0}(x), \qquad \frac{\partial }{\partial t}%
u(x,0)=v_{0}(x),
\end{equation}
with $f$ being continuous for $t\geq 0$, $u_{0}\in C^{1}(\mathbb{R})$ and $%
v_{0}\in C(\mathbb{R})$. Its classical solution has $C^{2}$ regularity for $%
t>0$, $C^{1}$ regularity for $t\geq 0$, satisfies equation \eqref{mp-cp} for 
$t>0$, and initial conditions as $t\to 0$. In the distributional setting $%
\mathcal{D} ^{\prime }(\mathbb{R}^2)$, if $u$ and $f$ are continued by zero
for $t<0$, equation 
\begin{equation}  \label{mp-gencp}
Pu=f(x,t) +u_{0}(x) \delta^{\prime }(t) +v_{0}(x) \delta (t)
\end{equation}
is satisfied, and is called the generalized Cauchy problem for the operator $%
P$. Then the classical solutions of \eqref{mp-cp} are among generalized
solutions of \eqref{mp-gencp} that vanish for $t<0$. If the operator $P$ has
a fundamental solution $E$, then 
\begin{equation*}
u=E\ast (f(x,t) +u_{0}(x) \delta^{\prime }(t) +v_{0}(x) \delta(t))
\end{equation*}
is a unique solution to the corresponding generalized Cauchy problem for $P$%
, see \cite{vlad}.


\section{Distributed order fractional wave equation}

\label{sec:dofzwe} 

We shall consider the Cauchy problem on infinite one-dimensional spatial
domain, i.e. problem on the real line $\mathbb{R}$, with $t>0$, for the
dimensionless system of equations 
\begin{eqnarray}
\frac{\partial }{\partial x}\sigma (x,t) &=&\frac{\partial ^{2}}{\partial
t^{2}}u(x,t),  \label{system-1} \\
\int_{0}^{1}\phi _{\sigma }(\alpha )\,{}_{0}D_{t}^{\alpha }\sigma
(x,t)\,d\alpha &=&\int_{0}^{1}\phi _{\varepsilon }(\alpha
)\,{}_{0}D_{t}^{\alpha }\varepsilon (x,t)\,d\alpha,  \label{system-2} \\
\varepsilon (x,t) &=&\frac{\partial }{\partial x}u(x,t),  \label{system-3}
\end{eqnarray}%
obtained from the initial system of equations (\ref{eq-mot}) - (\ref{strain}%
) after introducing dimensionless quantities: $\bar{x}=x/X^{\ast }$, $\bar{t}%
=t/T^{\ast }$, $\bar{u}=u/X^{\ast }$, $\bar{\sigma}=\sigma /E$, $\bar{%
\varepsilon}=\varepsilon $, $\bar{\phi}_{\sigma }=\phi _{\sigma }/(T^{\ast
})^{\alpha }$, $\bar{\phi}_{\varepsilon }=\phi _{\varepsilon }/(T^{\ast
})^{\alpha }$, with $X^{\ast }=T^{\ast }\sqrt{E/\rho }$ and $T^{\ast }$
determined differently for each particular constitutive model, for example
according to one of the generalized time constants in $\phi _{\varepsilon }$%
. We refer to \cite{KOZ10} for more details. Thus, we shall look for
solutions to system of equations \eqref{system-1} - \eqref{system-3} which
satisfy initial and boundary conditions: 
\begin{eqnarray}
&u(x,0)=u_{0}(x),\quad \displaystyle\frac{\partial }{\partial t}%
u(x,0)=v_{0}(x),\quad \sigma (x,0)=0,\quad \varepsilon (x,0)=0,&  \label{ic}
\\
&\lim_{x\rightarrow \pm \infty }u(x,t)=0,\qquad \lim_{x\rightarrow \pm
\infty }\sigma (x,t)=0.&  \label{bc}
\end{eqnarray}

More precisely, instead of examining system of equations \eqref{system-1} - %
\eqref{system-3}, we shall be concerned with the following equation: 
\begin{equation}
\frac{\partial ^{2}}{\partial t^{2}}u(x,t)-L(t)\frac{\partial ^{2}}{\partial
x^{2}}u(x,t)=0,\qquad x\in \mathbb{R},\,\,t>0,  \label{dowe}
\end{equation}%
with $L$ being a linear operator of convolution type acting on $\mathcal{S}%
^{\prime }(\mathbb{R})$, given by \eqref{opP}, that will be derived below.
It will be shown that, under certain assumptions on $\phi _{\varepsilon }$
and $\phi _{\sigma }$, system of equations \eqref{system-1} - %
\eqref{system-3} and equation \eqref{dowe} are equivalent. Equation %
\eqref{dowe} will be called the wave equation for distributed order type
viscoelastic media.

The first result that will be presented provides equivalence of system of
equations \eqref{system-1} - \eqref{system-3} and equation \eqref{dowe}, so
we shall specify regularity properties of functions that define our model.
We shall assume that the constitutive relation \eqref{system-2} is
determined by compactly supported distributions $\phi _{\sigma },\phi
_{\varepsilon }\in \mathcal{E}^{\prime }(\mathbb{R})$ with support in $[0,1]$%
, while the initial conditions $u_{0}$ and $v_{0}$ are assumed to be
elements of $\mathcal{S}^{\prime }(\mathbb{R})$.

\begin{thm}
\label{th:eq-s-we} Let $\phi _{\sigma },\phi _{\varepsilon }\in \mathcal{E}%
^{\prime }(\mathbb{R})$ with support in $[0,1]$. Set 
\begin{equation}
\Phi _{\sigma }(s)=\langle \phi _{\sigma }(\alpha ),s^{\alpha }\rangle \quad %
\mbox{ and }\quad \Phi _{\varepsilon }(s)=\langle \phi _{\varepsilon
}(\alpha ),s^{\alpha }\rangle ,\quad \mathop{\rm Re}\nolimits s>0.
\label{fiovi}
\end{equation}%
Suppose that the following assumption holds.

\begin{itemize}
\item \normalfont{{\bf{Assumption (A1):\ }}} $\displaystyle \mathcal{L}
^{-1} \bigg(\frac{\Phi_\varepsilon(s)}{\Phi_\sigma(s)}\bigg)$ exists as an
element of $\mathcal{S} ^{\prime }_+(\mathbb{R})$.
\end{itemize}

Then system of equations \eqref{system-1} - \eqref{system-3} and equation %
\eqref{dowe} with $L(t):= \mathcal{L} ^{-1} \Big(\frac{\Phi_\varepsilon(s)}{%
\Phi_\sigma(s)}\Big)\ast_{t}$, are equivalent.
\end{thm}

\noindent \textbf{Proof. } One first applies the Laplace transform with
respect to $t$ to \eqref{system-2} to obtain 
\begin{equation*}
\tilde{\sigma}(x,s)\langle \phi _{\sigma }(\alpha ),s^{\alpha }\rangle =%
\tilde{\varepsilon}(x,s)\langle \phi _{\varepsilon }(\alpha ),s^{\alpha
}\rangle .
\end{equation*}%
According to the assumption \textbf{(A1)} it follows that 
\begin{equation}
\sigma =\mathcal{L}^{-1}\bigg(\frac{\Phi _{\varepsilon }(s)}{\Phi _{\sigma
}(s)}\bigg)\ast _{t}\varepsilon  \label{solsigma}
\end{equation}%
is well-defined. Now, one inserts $\varepsilon $ from \eqref{system-3} into %
\eqref{solsigma}, as well as $\sigma $ from \eqref{solsigma} into %
\eqref{system-1}. As the result one obtains exactly \eqref{dowe}, and hence
equivalence of system \eqref{system-1} - \eqref{system-3} with equation %
\eqref{dowe} follows. \nolinebreak {\hspace*{\fill}$\Box $ \vspace*{0.25cm}}


Defining the operator 
\begin{equation}  \label{opP}
P:=\frac{\partial^{2}}{\partial t^{2}}- L(t) \frac{\partial^{2}}{\partial
x^{2}}, \qquad L(t):= \mathcal{L} ^{-1} \bigg(\frac{\Phi_\varepsilon(s)}{%
\Phi_\sigma(s)}\bigg)\ast_{t}
\end{equation}
we can write \eqref{dowe} in the form $Pu=0$.

In the sequel we shall study equation \eqref{dowe} in the distributional
setting. In fact, we shall seek a fundamental solution to the generalized
Cauchy problem for \eqref{opP} in $\mathcal{S}^{\prime }(\mathbb{R}\times 
\mathbb{R}_{+})$. This will also lead to solutions of the problem %
\eqref{system-1} - \eqref{bc}, due to equivalence of \eqref{system-1} - %
\eqref{system-3} and \eqref{dowe}.


\subsection{Solvability of the generalized Cauchy problem}

\label{sec:fund sol} 

The generalized Cauchy problem for the operator $P$ given by \eqref{opP}
takes the following form $Pu(x,t)=u_{0}(x)\delta^{\prime }(t)+v_{0}(x)\delta
(t)$, or equivalently 
\begin{equation}  \label{dowe-gCp}
\frac{\partial^{2}}{\partial t^{2}}u(x,t) = \mathcal{L} ^{-1} \bigg(\frac{%
\Phi_\varepsilon(s)}{\Phi_\sigma(s)}\bigg) \ast_{t} \frac{\partial^{2}}{%
\partial x^{2}}u(x,t) + u_{0}(x) \delta^{\prime }(t)+v_{0}(x)\delta (t).
\end{equation}
Clearly, initial conditions are included into the generalized Cauchy problem
(see Section \ref{sec:set&tol}), and we shall look for solutions that also
satisfy boundary conditions \eqref{bc}.

The theorem which follows provides conditions that guarantee existence and
uniqueness of a solution to the generalized Cauchy problem \eqref{dowe-gCp}.

\begin{thm}
\label{th:glavna} Let $u_{0},v_{0}\in \mathcal{S}^{\prime }(\mathbb{R})$.
Suppose that the assumption \textbf{(A1)} holds. Further, assume the
following.

\begin{itemize}
\item \normalfont{\textbf{Assumption (A2):\ }} $\displaystyle s^2 \frac{%
\Phi_\sigma(s)}{\Phi_\varepsilon(s)} \in \mathbb{C}\setminus (-\infty,0]$,
for all $s\in\mathbb{C}$ with $\mathop{\rm Re}\nolimits s>0$.

\item \normalfont{\textbf{Assumption (A3):\ }} $\displaystyle \mathcal{L}
^{-1} \bigg(\frac{\Phi_\sigma(s)}{\Phi_\varepsilon(s)}\bigg)$ exists as an
element of $\mathcal{S} ^{\prime }_+$.
\end{itemize}

Then there exists a unique solution $u\in\mathcal{S} ^{\prime }(\mathbb{R}%
\times \mathbb{R}_{+})$ to \eqref{dowe-gCp} given by 
\begin{equation}  \label{sol-u}
u(x,t)=S(x,t) \ast_{x,t}(u_{0}(x)\delta^{\prime }(t)+v_{0}(x)\delta(t)),
\end{equation}
where $S\in\mathcal{S} ^{\prime }(\mathbb{R}\times\mathbb{R}_{+})$ is a
fundamental solution of the operator $P$.
\end{thm}

\noindent \textbf{Proof. } We first apply the Laplace transform with respect
to $t$ to \eqref{dowe-gCp} and obtain 
\begin{equation}
\frac{\partial ^{2}}{\partial x^{2}}\tilde{u}(x,s)-s^{2}\frac{\Phi _{\sigma
}(s)}{\Phi _{\varepsilon }(s)}\tilde{u}(x,s)=-\frac{\Phi _{\sigma }(s)}{\Phi
_{\varepsilon }(s)}\big(su_{0}(x)+v_{0}(x)\big).  \label{lt-fzwe-distr}
\end{equation}%
Then \eqref{lt-fzwe-distr} is of the form 
\begin{equation}
v^{\prime \prime }-\omega v=-f\quad \mbox{ with }\quad \omega (s):=s^{2}%
\frac{\Phi _{\sigma }(s)}{\Phi _{\varepsilon }(s)},\,\,f(x,s):=\frac{\Phi
_{\sigma }(s)}{\Phi _{\varepsilon }(s)}(su_{0}(x)+v_{0}(x)),\,\,\mathop{\rm
Re}\nolimits s>0.  \label{lem-fZwe}
\end{equation}%
The latter type of equation was studied in details in \cite{KOZ10}, where it
was proved that its solution takes the form $v=\frac{e^{-\sqrt{\omega }|x|}}{%
2\sqrt{\omega }}\ast f$, if $f(\cdot ,s)\in \mathcal{S}^{\prime }(\mathbb{R}%
) $ and $\omega \in \mathbb{C}\setminus (-\infty ,0]$. In the present
situation, the first condition on $f$ is clearly satisfied, since by
assumption $u_{0},v_{0}\in \mathcal{S}^{\prime }(\mathbb{R})$, while the
second condition on $\omega $ is exactly the assumption \textbf{(A2)}.

Therefore, the corresponding solution to \eqref{lt-fzwe-distr} reads: 
\begin{eqnarray}
\tilde{u}(x,s) &=& \frac{e^{-\sqrt{\omega(s)} |x|}}{2\sqrt{\omega(s)}}
\ast_{x} \frac{\omega (s)}{s^{2}}(su_{0}(x)+v_{0}(x))  \notag \\
&=& \frac{\sqrt{\omega (s)}e^{-\sqrt{\omega(s)} |x|}}{2s^{2}} \ast_{x}
(su_{0}(x)+v_{0}(x))  \notag \\
&=& \frac{1}{2s}\sqrt{\frac{\Phi_\sigma(s)}{\Phi_\varepsilon(s)}} e^{-|x|s%
\sqrt{\frac{\Phi_\sigma(s)}{\Phi_\varepsilon(s)}}} \ast_{x}
(su_{0}(x)+v_{0}(x)).  \label{tilde u}
\end{eqnarray}
Solution $u$ to \eqref{dowe} is obtained by applying the inverse Laplace
transform to \eqref{tilde u}. Set 
\begin{equation}  \label{tildeS}
\tilde{S}(x,s):=\frac{1}{2s}\sqrt{\frac{\Phi_\sigma(s)}{\Phi_\varepsilon(s)}}
e^{-|x|s\sqrt{\frac{\Phi_\sigma(s)}{\Phi_\varepsilon(s)}}}, \qquad x\in 
\mathbb{R},\,\, \mathop{\rm Re}\nolimits s>0,
\end{equation}
and $S(x,t): =\mathcal{L} ^{-1}(\tilde{S}(x,s))(t)$, $x\in\mathbb{R}$, $t>0$%
. Existence of the inverse Laplace transform of $\tilde{S}$ follows from the
assumptions \textbf{(A2)} and \textbf{(A3)}. Indeed, $\mathop{\rm Re}%
\nolimits \Big(s\sqrt{\frac{\Phi_\sigma (s)}{\Phi_\varepsilon (s)}}\Big)>0$,
which follows from \textbf{(A2)}, while from \textbf{(A3)} it follows that
for all $x\in\mathbb{R}$ and $s\in\mathbb{C}_+$ there exist constants $a,b\in%
\mathbb{R}$ and $C_1>0$ such that $\Big| \frac{\Phi_\sigma (s)}{%
\Phi_\varepsilon (s)} \Big|
\leq C_1\frac{(1+|s|)^a}{(\mathop{\rm Re}\nolimits s )^b}$. Therefore, $\Big|%
\displaystyle \frac{1}{2s}\sqrt{\frac{\Phi_\sigma (s)}{\Phi_\varepsilon (s)}}
e^{-|x|s\sqrt{\frac{\Phi_\sigma (s)}{\Phi_\varepsilon (s)}}}\Big|\leq C 
\frac{(1+|s|)^m}{(\mathop{\rm Re}\nolimits s )^k} e^{-|x|\mathop{\rm Re}%
\nolimits \Big(s\sqrt{\frac{\Phi_\sigma (s)}{\Phi_\varepsilon (s)}}\Big)}
\leq C \frac{(1+|s|)^m}{(\mathop{\rm Re}\nolimits s )^k}$, for some $m,k\in%
\mathbb{R}$ and $C>0$. Such $S$ is a fundamental solution of the operator $P 
$, and it is well-defined element in $\mathcal{S} ^{\prime }_+(\mathbb{R})$.
Hence, solution $u$ to \eqref{dowe-gCp} is given by \eqref{sol-u}, which
proves the claim. \nolinebreak{\hspace*{\fill}$\Box$ \vspace*{0.25cm}}

\begin{rem}
\textrm{\label{rem:vvzn} }

\begin{itemize}
\item[(i)] Note that the assumptions \textbf{(A1)} and \textbf{(A3)}
guarantee that the distributed order constitutive equation (\ref{system-2})
can be solved for both stress and strain, while the assumption \textbf{(A2)}
ensures the decay of fundamental solution (\ref{tildeS}) to zero at
infinity, since $\mathop{\rm Re}\nolimits\left( s\sqrt{\frac{\Phi _{\sigma
}(s)}{\Phi _{\varepsilon }(s)}}\right) >0$, for $\mathop{\rm Re}\nolimits
s>0 $.

\item[(ii)] From \eqref{tilde u} we further have: 
\begin{eqnarray}
\tilde{u}(x,s) &=&\frac{1}{2s}\sqrt{\frac{\Phi _{\sigma }(s)}{\Phi
_{\varepsilon }(s)}}e^{-|x|s\sqrt{\frac{\Phi _{\sigma }(s)}{\Phi
_{\varepsilon }(s)}}}\ast _{x}(su_{0}(x)+v_{0}(x))  \notag \\
&=&\frac{1}{2}\sqrt{\frac{\Phi _{\sigma }(s)}{\Phi _{\varepsilon }(s)}}%
e^{-|x|s\sqrt{\frac{\Phi _{\sigma }(s)}{\Phi _{\varepsilon }(s)}}}\ast _{x}%
\Big(u_{0}(x)+\frac{1}{s}v_{0}(x)\Big).  \label{u-tilda-S'}
\end{eqnarray}%
Thus, instead of $\tilde{S}$ defined by \eqref{tildeS} we can consider 
\begin{equation}
\tilde{K}(x,s):=\frac{1}{2}\sqrt{\frac{\Phi _{\sigma }(s)}{\Phi
_{\varepsilon }(s)}}e^{-|x|s\sqrt{\frac{\Phi _{\sigma }(s)}{\Phi
_{\varepsilon }(s)}}},\qquad x\in \mathbb{R},\,\,\mathop{\rm Re}\nolimits
s>0,  \label{tildeS'}
\end{equation}%
and $K(x,t):=\mathcal{L}^{-1}(\tilde{K}(x,s))(t)$, $x\in \mathbb{R}$, $t>0$.
We shall refer to both $S$ and $K$ as the fundamental solution.

\item[(iii)] In some situations, as it will become apparent in the sequel,
it is more convenient to use an alternative form of the assumption \textbf{%
(A2)} from Theorem \ref{th:glavna}, that is given by

\begin{itemize}
\item 
\normalfont{{\textbf{Assumption (A2'):\ }} $\psi (\xi ,s):=s^{2}+\xi ^{2}\frac{\Phi _{\varepsilon }(s)}{\Phi _{\sigma }(s)}\not=0$, for all $\xi \in 
\mathbb{R}$ and $s\in \mathbb{C}$ with $\mathop{\rm Re}\nolimits s>0$. }
\end{itemize}

Equivalence of \textbf{(A2)} and \textbf{(A2')} is straight forward. For $%
s\in \mathbb{C}$ with $\mathop{\rm Re}\nolimits s>0$, and $\xi \in \mathbb{R}
$, 
\begin{equation*}
s^{2}\frac{\Phi _{\sigma }(s)}{\Phi _{\varepsilon }(s)}\in \mathbb{C}%
\setminus (-\infty ,0]\quad \Leftrightarrow \quad s^{2}\frac{\Phi _{\sigma
}(s)}{\Phi _{\varepsilon }(s)}+\xi ^{2}\not=0\quad \Leftrightarrow \quad
s^{2}+\xi ^{2}\frac{\Phi _{\varepsilon }(s)}{\Phi _{\sigma }(s)}\not=0.
\end{equation*}

This can also be seen from the proof of Theorem \ref{th:glavna}. By applying
the Fourier transform to \eqref{lt-fzwe-distr}, one obtains 
\begin{equation}
-\xi ^{2}\hat{\tilde{u}}(\xi ,s)-s^{2}\frac{\Phi _{\sigma }(s)}{\Phi
_{\varepsilon }(s)}\hat{\tilde{u}}(\xi ,s)=-\frac{\Phi _{\sigma }(s)}{\Phi
_{\varepsilon }(s)}\big(s\hat{u}_{0}(\xi )+\hat{v}_{0}(\xi )\big),
\label{mdmn}
\end{equation}%
or equivalently 
\begin{equation*}
\hat{\tilde{u}}(\xi ,s)=\frac{s\hat{u}_{0}(\xi )+\hat{v}_{0}(\xi )}{%
s^{2}+\xi ^{2}\frac{\Phi _{\varepsilon }(s)}{\Phi _{\sigma }(s)}},\qquad \xi
\in \mathbb{R},\,\,\mathop{\rm Re}\nolimits s>0,
\end{equation*}%
and hence, assumption \textbf{(A2')} provides solvability of \eqref{mdmn}.

\item[(iv)] Notice that from \eqref{tildeS} one has that the fundamental
solution vanishes as $x\rightarrow \pm \infty $, therefore the unique
solution \eqref{sol-u} to the generalized Cauchy problem satisfies boundary
conditions \eqref{bc}.

\item[(v)] Nonhomogeneous case of \eqref{dowe}, i.e., a rod under the
influence of body forces, can be analogously treated, cf.\ \cite{KOZ10}.
\end{itemize}
\end{rem}

Once the existence and uniqueness of a solution to the generalized Cauchy
problem is proved, one immediately has the following corollary.

\begin{cor}
Suppose that the assumptions \textbf{(A1)}-\textbf{(A3)} hold. Let $u$ be
given by \eqref{sol-u}. Then 
\begin{equation*}
(u,\varepsilon ,\sigma )(x,t)=\bigg(u(x,t),\frac{\partial }{\partial x}%
u(x,t),\mathcal{L}^{-1}\bigg(\frac{\Phi _{\varepsilon }(s)}{\Phi _{\sigma
}(s)}\bigg)\ast _{t}\frac{\partial }{\partial x}u(x,t)\bigg)\in (\mathcal{S}%
^{\prime }(\mathbb{R}\times \mathbb{R}_{+}))^{3},
\end{equation*}%
is a unique solution to the system \eqref{system-1} - \eqref{bc}.
\end{cor}


\subsection{Calculation of the fundamental solution}

\label{sec:calculsol} 

In Theorem \ref{th:glavna} we proved the existence of inverse Laplace
transform of the fundamental solution \eqref{tildeS} and obtained solution $u
$ as the convolution of fundamental solution $S$ with initial conditions.
Closer inspection of the proof of theorem indicates that in order to
calculate $S$ explicitly one needs to impose additional assumptions. This is
considered in the following statement.

\begin{thm}
\label{th:racunaljka} Suppose that the assumptions \textbf{(A1)}, \textbf{%
(A2)} and \textbf{(A3)} of Theorems \ref{th:eq-s-we} and \ref{th:glavna}
hold. Suppose in addition the following.

\begin{itemize}
\item \normalfont{\textbf{Assumption (A4):\ }} Multiform function $\tilde{K}$
given by \eqref{tildeS'}, has only two branch points $s=0$ and $s=\infty $.

\item \normalfont{\textbf{Assumption (A5):\ }} $\displaystyle%
\lim_{R\rightarrow \infty }\bigg|\sqrt{\frac{\Phi _{\sigma }(Re^{i\varphi })%
}{\Phi _{\varepsilon }(Re^{i\varphi })}}\bigg|=k$, for $\displaystyle\varphi
\in \Big(\frac{\pi }{2},\pi \Big)\cup \Big(-\pi ,-\frac{\pi }{2}\Big)$ and $%
k\geq 0$.

\item \normalfont{\textbf{Assumption (A6):\ }} $\displaystyle \lim_{\eta\to
0} \bigg|
\eta \sqrt{\frac{\Phi_\sigma (\eta e^{i\varphi})}{\Phi_\varepsilon (\eta
e^{i\varphi})}} \bigg| =0$, for $\varphi\in(-\pi,\pi)$.
\end{itemize}

Then the solution $u$ reads 
\begin{equation}
u(x,t)=K(x,t)\ast _{x,t}(u_{0}(x)\delta (t)+v_{0}(x)H(t)),
\label{fund sol u}
\end{equation}%
where the fundamental solution $K$ can be calculated as 
\begin{equation}
K(x,t)=\frac{1}{4\pi i}\int_{0}^{\infty }\Bigg(\sqrt{\frac{\Phi _{\sigma
}(qe^{-i\pi })}{\Phi _{\varepsilon }(qe^{-i\pi })}}e^{|x|q\sqrt{\frac{\Phi
_{\sigma }(qe^{-i\pi })}{\Phi _{\varepsilon }(qe^{-i\pi })}}}-\sqrt{\frac{%
\Phi _{\sigma }(qe^{i\pi })}{\Phi _{\varepsilon }(qe^{i\pi })}}e^{|x|q\sqrt{%
\frac{\Phi _{\sigma }(qe^{i\pi })}{\Phi _{\varepsilon }(qe^{i\pi })}}}\Bigg)%
e^{-qt}\,dq,  \label{fund sol}
\end{equation}%
$K\in \mathcal{S}^{\prime }(\mathbb{R}\times \mathbb{R}_{+})$, and has the
support in the cone $|x|<ct$, with $c=1/k$. If $|x| > ct,$ then $K=0$.
\end{thm}

\noindent \textbf{Proof. }The form of solution \eqref{fund sol u} follows
from \eqref{u-tilda-S'}. Due to the assumption \textbf{(A4)}, the inverse
Laplace transform of multiform function $\tilde{K}$, given by \eqref{tildeS'}%
, can be calculated by the use of the Cauchy integral theorem: 
\begin{equation}
\oint_{\Gamma }\tilde{K}(x,s)e^{st}\,ds=0,\qquad x\in \mathbb{R},\,\,t>0.
\label{Cauchy int th}
\end{equation}%
Here $\Gamma =\Gamma _{1}\cup \Gamma _{2}\cup \Gamma _{\eta }\cup \Gamma
_{3}\cup \Gamma _{4}\cup \gamma _{0}$ is a contour given by

\noindent%
\parbox{0.55\textwidth}{
 $$
 \begin{array}{rl}
 \Gamma_{1}: & s=Re^{i\vphi},\vphi_{0}<\vphi <\pi; \\
 \Gamma_{2}: & s=qe^{i\pi},-R<-q<-\eta; \\
 \Gamma_{\eta}: & s=\eta e^{i\vphi},-\pi <\vphi <\pi; \\
 \Gamma_{3}: & s=qe^{-i\pi},\eta <q<R; \\
 \Gamma_{4}: & s=Re^{i\vphi},-\pi <\vphi<-\vphi_{0}; \\
 \gamma_{0}: & s=a(1+i\tan \vphi),-\vphi_{0}<\vphi <\vphi_{0},
 \end{array}
 $$
 } \hphantom{Bla} 
\parbox{0.35\textwidth}{
 \includegraphics[width=5cm]{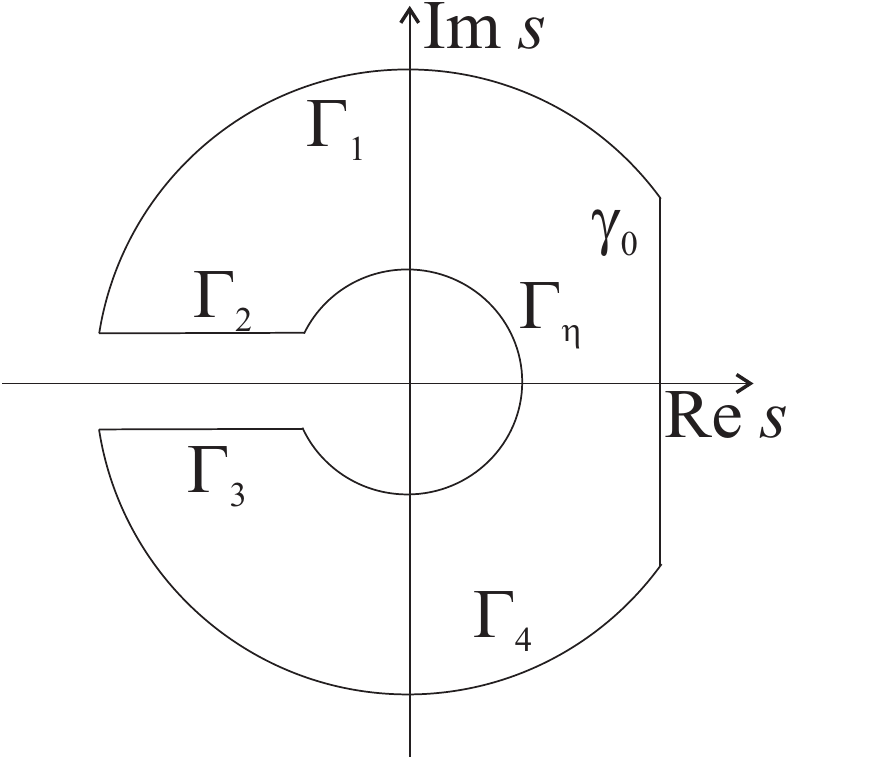}
 }

\noindent where $R>0$, $0<\eta<R$, $a>0$ and $\varphi_{0}=\arccos(\frac{a}{R}%
)$. Note that $\lim_{R\to\infty} \varphi_{0}=\frac{\pi}{2}$.

In the limit when $R\rightarrow \infty $, integral along contour $\Gamma
_{1} $ reads ($x\in \mathbb{R},t>0$) 
\begin{equation*}
\lim_{R\rightarrow \infty }\int_{\Gamma _{1}}\tilde{K}(x,s)e^{st}\,ds=\frac{1%
}{2}\lim_{R\rightarrow \infty }\int_{\varphi _{0}}^{\pi }\sqrt{\frac{\Phi
_{\sigma }(Re^{i\varphi })}{\Phi _{\varepsilon }(Re^{i\varphi })}}%
e^{-|x|Re^{i\varphi }\sqrt{\frac{\Phi _{\sigma }(Re^{i\varphi })}{\Phi
_{\varepsilon }(Re^{i\varphi })}}+Rte^{i\varphi }}Rie^{i\varphi }\,d\varphi .
\end{equation*}%
Thus, we have 
\begin{eqnarray*}
\lim_{R\rightarrow \infty }\bigg|\int_{\Gamma _{1}}\tilde{K}(x,s)e^{st}\,ds%
\bigg| &\leq &\frac{1}{2}\lim_{R\rightarrow \infty }\int_{\varphi _{0}}^{\pi
}\bigg|\sqrt{\frac{\Phi _{\sigma }(Re^{i\varphi })}{\Phi _{\varepsilon
}(Re^{i\varphi })}}\bigg|\bigg|e^{-|x|Re^{i\varphi }\sqrt{\frac{\Phi
_{\sigma }(Re^{i\varphi })}{\Phi _{\varepsilon }(Re^{i\varphi })}}}\bigg|%
e^{Rt\cos \varphi }R\,d\varphi \\
&\leq &\frac{k}{2}\lim_{R\rightarrow \infty }\int_{\varphi _{0}}^{\pi
}e^{R\cos \varphi (t-k|x|)}R\,d\varphi =0,\quad \mbox{ if }\quad t>k|x|,
\end{eqnarray*}%
where the last inequality follows from the assumption \textbf{(A5)}, while
the equality is due to the fact that $\cos \varphi <0$ for $\varphi \in (%
\frac{\pi }{2},\pi )$. Notice that in the case $k=0$ the integral along $%
\Gamma _{1}$ vanishes as $R\rightarrow \infty $ for all $x\in \mathbb{R}$
and $t>0$.

Similar argument is valid for the integral along $\Gamma_{4}$.

Integral along $\Gamma _{\eta }$ also tends to zero when $\eta \rightarrow 0$%
. To see this we use the assumption \textbf{(A6)} to obtain 
\begin{equation*}
\lim_{\eta \rightarrow 0}\bigg|\int_{\Gamma _{\eta }}\tilde{K}(x,s)e^{st}\,ds%
\bigg|\leq \frac{1}{2}\lim_{\eta \rightarrow 0}\int_{\pi }^{-\pi }\bigg|\eta 
\sqrt{\frac{\Phi _{\sigma }(\eta e^{i\varphi })}{\Phi _{\varepsilon }(\eta
e^{i\varphi })}}\bigg|\bigg|e^{-|x|\eta e^{i\varphi }\sqrt{\frac{\Phi
_{\sigma }(\eta e^{i\varphi })}{\Phi _{\varepsilon }(\eta e^{i\varphi })}}}%
\bigg|e^{\eta t\cos \varphi }\,d\varphi =0.
\end{equation*}

In the limit when $R\rightarrow \infty $ and $\eta \rightarrow 0$, integrals
along contours $\Gamma _{2}$, $\Gamma _{3}$ and $\gamma _{0}$ read: 
\begin{eqnarray*}
\lim_{\substack{ R\rightarrow \infty  \\ \eta \rightarrow 0}}\int_{\Gamma
_{2}}\tilde{K}(x,s)e^{st}\,ds &=&\frac{1}{2}\int_{0}^{\infty }\sqrt{\frac{%
\Phi _{\sigma }(qe^{i\pi })}{\Phi _{\varepsilon }(qe^{i\pi })}}e^{-q\Big(%
t-|x|\sqrt{\frac{\Phi _{\sigma }(qe^{i\pi })}{\Phi _{\varepsilon }(qe^{i\pi
})}}\Big)}\,dq \\
\lim_{\substack{ R\rightarrow \infty  \\ \eta \rightarrow 0}}\int_{\Gamma
_{3}}\tilde{K}(x,s)e^{st}\,ds &=&-\frac{1}{2}\int_{0}^{\infty }\sqrt{\frac{%
\Phi _{\sigma }(qe^{-i\pi })}{\Phi _{\varepsilon }(qe^{-i\pi })}}e^{-q\Big(%
t-|x|\sqrt{\frac{\Phi _{\sigma }(qe^{-i\pi })}{\Phi _{\varepsilon
}(qe^{-i\pi })}}\Big)}\,dq \\
\lim_{R\rightarrow \infty }\int_{\gamma _{0}}\tilde{K}(x,s)e^{st}\,ds
&=&2\pi iK(x,t).
\end{eqnarray*}%
Thus, it follows from the Cauchy integral theorem \eqref{Cauchy int th} that 
$K$ takes the form as in \eqref{fund sol}.

In the domain $|x|>ct$, $t>0$, function $K$ can be calculated by using the
Cauchy integral formula 
\begin{equation}
\oint_{\bar{\Gamma}}\tilde{K}(x,s)e^{st}\,ds=0,\qquad x\in \mathbb{R}%
,\,\,t>0,  \label{eq:Cauchy int for-cb-1}
\end{equation}%
where $\bar{\Gamma}=\bar{\Gamma}_{1}\cup \gamma _{0}$ is the contour
parametrized by

\noindent%
\parbox{0.55\textwidth}{
 $$
 \begin{array}{rl}
  \bar{\Gamma}_{1} &: s=a+Re^{i\vphi}, \,\, -\frac{\pi}{2}<\vphi <\frac{\pi}{2}; \\
  \gamma_{0} &: s=a(1+i\tg \vphi ), \,\, -\vphi_{0}<\vphi <\vphi_{0},
 \end{array}
 $$
 } \hphantom{Bla} 
\parbox{0.35\textwidth}{
 \includegraphics[width=3cm]{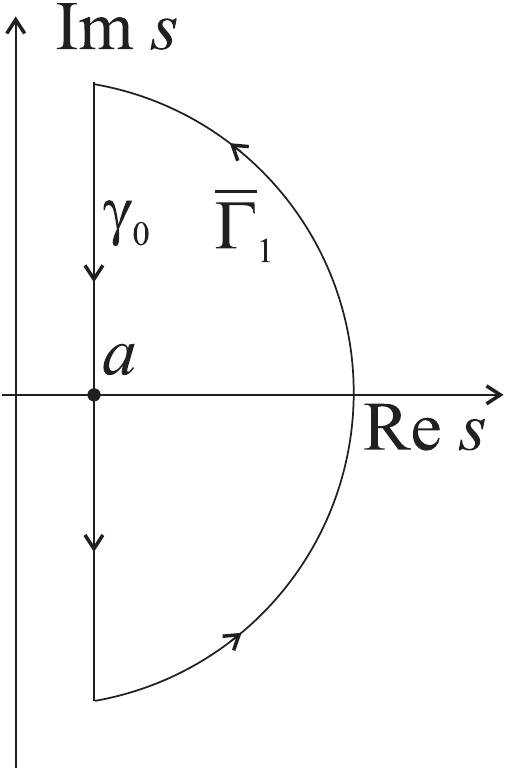}
 }

\noindent with $\varphi _{0}=\mathop{\rm arctg}\nolimits\frac{R}{a}\in (-%
\frac{\pi }{2},\frac{\pi }{2})$. In the limit when $R\rightarrow \infty $,
integral along the contour $\bar{\Gamma}_{1}$ reads 
\begin{eqnarray*}
\lim_{R\rightarrow \infty }\bigg|\int_{\bar{\Gamma}_{1}}\tilde{K}%
(x,s)e^{st}\,ds\bigg| &=&\frac{1}{2}\lim_{R\rightarrow \infty
}\int_{-\varphi _{0}}^{\varphi _{0}}\bigg|\sqrt{\frac{\Phi _{\sigma
}(Re^{i\varphi })}{\Phi _{\varepsilon }(Re^{i\varphi })}}\bigg|\bigg|%
e^{-|x|Re^{i\varphi }\sqrt{\frac{\Phi _{\sigma }(Re^{i\varphi })}{\Phi
_{\varepsilon }(Re^{i\varphi })}}}\bigg|e^{Rt\cos \varphi }R\,d\varphi \\
&\leq &\frac{K}{2}\lim_{R\rightarrow \infty }\int_{-\varphi _{0}}^{\varphi
_{0}}e^{-R\cos \varphi (k|x|-t)}R\,d\varphi =0,\quad \mbox{ if }\quad t<k|x|,
\end{eqnarray*}%
due to the assumption \textbf{(A5)} and the fact that $\cos \varphi >0$, for 
$\varphi \in (-\frac{\pi }{2},\frac{\pi }{2})$. Thus, by 
\eqref{eq:Cauchy
int for-cb-1} we have $K(x,t)=0$, for $|x|>ct$. \nolinebreak {\hspace*{\fill}%
$\Box $ \vspace*{0.25cm}}


Assumption (\textbf{A5}) introduces the constant $k=1/c$, which, for
non-zero values of $k$, defines the domain of fundamental solution $K$ as
the cone $|x|<ct$, outside which $K=0.$ This implies that the constant $c$
can be interpreted as the (dimensionless) wave propagation speed in
viscoelastic media described by distributed order fractional model. Turning
to the dimensional coordinate and time, the wave propagation speed is $%
c_{vm}=c\sqrt{\frac{E}{\rho }}$.

The wave propagation speed is closely related to the material properties in
creep and stress relaxation through the assumption (\textbf{A5}). Namely,
the constitutive equation \eqref{system-2}, solved with respect to stress is
given by \eqref{solsigma}, and its Laplace transform is 
\begin{equation*}
\tilde{\sigma}(s)=\frac{\Phi _{\varepsilon }(s)}{\Phi _{\sigma }(s)}\tilde{%
\varepsilon}(s).
\end{equation*}%
The creep compliance, i.e., strain in the creep experiment (stress is
assumed to be the Heaviside function), is 
\begin{equation}
\tilde{J}(s)=\frac{1}{s}\frac{\Phi _{\sigma }(s)}{\Phi _{\varepsilon }(s)}.
\label{J-tilda}
\end{equation}%
Then, the glass compliance is 
\begin{equation*}
J_{g}:=\lim_{t\rightarrow 0}J\left( t\right) =\lim_{s\rightarrow \infty }(s%
\tilde{J}(s))=\lim_{s\rightarrow \infty }\frac{\Phi _{\sigma }(s)}{\Phi
_{\varepsilon }(s)}=k^{2}=\frac{1}{c^{2}},
\end{equation*}%
with $k$ from assumption (\textbf{A5}) and $J(t)=\mathcal{L}^{-1}\big(\tilde{%
J}(s)\big)(t)$. The relaxation modulus, i.e., stress in the stress
relaxation experiment (strain is assumed to be the Heaviside function), is
connected to the creep compliance by 
\begin{equation*}
s\tilde{G}(s)=\frac{1}{s\tilde{J}(s)}=\frac{\Phi _{\varepsilon }(s)}{\Phi
_{\sigma }(s)},
\end{equation*}%
so that the glass modulus is 
\begin{equation}
G_{g}:=\lim_{t\rightarrow 0}G\left( t\right) =\lim_{s\rightarrow \infty }(s%
\tilde{G}(s))=\frac{1}{J_{g}}=\lim_{s\rightarrow \infty }\frac{\Phi
_{\varepsilon }(s)}{\Phi _{\sigma }(s)},  \label{gege}
\end{equation}%
where $G(t)=\mathcal{L}^{-1}\big(\tilde{G}(s)\big)(t)$. Hence, the wave
speed in the distributed order fractional viscoelastic media is obtained as 
\begin{equation}
c=\sqrt{G_{g}}=\frac{1}{\sqrt{J_{g}}},  \label{ce}
\end{equation}%
if the glass modulus (compliance) is finite (non-zero), i.e., the wave speed
is determined by the finite initial value of the stress (strain) in the
stress relaxation (creep) experiment. If the glass modulus (compliance) is
infinite (zero), then we only conclude that the fundamental solution takes
the form \eqref{fund sol} for all $x\in \mathbb{R}$, and $t>0$, without a
straightforward indication about the wave speed through the solution support
properties. This case requires other tools in analyzing solution properties.
The same result, obtained by considering the wave-propagation problem on
semi-infinite domain, without detailed explanation is reported in \cite%
{Mai-10}.

Consider the assumption (\textbf{A6}) rewritten as 
\begin{equation*}
\lim_{s\rightarrow 0}\left( s\sqrt{s\tilde{J}(s)}\right) =\lim_{s\rightarrow
0}\left( \sqrt{s^{3}\tilde{J}(s)}\right) =0
\end{equation*}%
using (\ref{J-tilda}). In order to have the previous limit satisfied, one
imposes the behavior of the creep compliance as%
\begin{equation*}
\tilde{J}(s)\sim \frac{K}{s^{3-\varepsilon }},\;\;\text{as}\;\;s\rightarrow
0,\;\;\text{for}\;\;\varepsilon >0,
\end{equation*}%
which, using the Tauberian theorem \cite[p. 498, Satz 3.]{dec-band1}, implies%
\begin{equation*}
J(t)\sim K\frac{t^{2-\varepsilon }}{\Gamma (3-\varepsilon )},\;\;\text{as}%
\;\;t\rightarrow \infty .
\end{equation*}%
Thus, assumption (\textbf{A6}) restricts the growth of the creep compliance
for large times.


\section{Linear fractional model}

\label{sec:lfmS} 

Linear fractional model of viscoelastic body 
\begin{equation}
\sum_{i=1}^{n}a_{i}\,{}_{0}D_{t}^{\alpha _{i}}\sigma
(x,t)=\sum_{j=1}^{m}b_{i}\,{}_{0}D_{t}^{\beta _{j}}\varepsilon (x,t),
\label{gen-lin}
\end{equation}%
involving the Riemann-Liouville fractional derivatives of orders belonging
to interval $[0,1)$, is a special case of the distributed order constitutive
stress-strain relation \eqref{system-2}. Considering thermodynamically
consistent linear fractional models, analyzed in \cite{AKOZ}, it can be
shown that the assumptions \textbf{(A1)}-\textbf{(A6)} are satisfied,
implying the existence and explicit form of a solution to the whole class of
fractional wave equations arising from these models. Moreover, each case of
thermodynamically consistent model corresponds to a known class of linear
constitutive equations of viscoelastic body, as presented in \cite{Mai-10}.
In the sequel, we discuss the cases of thermodynamically consistent linear
fractional models of viscoelastic body and also relate the obtained results
to the previous research.

In \cite{AKOZ}, we studied thermodynamical restrictions for the most general
linear fractional constitutive equations (\ref{gen-lin}), and described
classes of admissible linear fractional models. The results of that
investigation can be summarized as follows.

\begin{itemize}
\item The highest order of fractional derivatives of stress cannot be
greater than the highest order of fractional derivatives of strain.

\item There are four admissible cases with respect to the orders of
fractional derivatives of stress and strain:

\begin{itemize}
\item[Case 1.] 
\begin{equation}  \label{case1}
\phi_\sigma(\alpha):=\sum_{i=1}^{n} a_{i} \delta(\alpha-\alpha_i), \quad
\phi_\varepsilon(\alpha):=\sum_{i=1}^{n} b_{i} \delta(\alpha-\alpha_i),
\end{equation}
with $0\leq\alpha_1<\ldots<\alpha_n< 1$, and $\frac{a_1}{b_1}\geq\frac{a_2}{%
b_2}\geq\ldots\geq\frac{a_n}{b_n}\geq 0$;

\item[Case 2.] 
\begin{equation}  \label{case2}
\phi_\sigma(\alpha):=\sum_{i=1}^{n} a_{i} \delta(\alpha-\alpha_i), \quad
\phi_\varepsilon(\alpha):=\sum_{i=1}^{n} b_{i} \delta(\alpha-\alpha_i) +
\sum_{i=n+1}^m b_{i} \delta(\alpha-\beta_i),
\end{equation}
with $0\leq\alpha_1<\ldots<\alpha_n<\beta_{n+1}<\ldots<\beta_m<1$, and $%
\frac{a_1}{b_1}\geq\frac{a_2}{b_2}\geq\ldots\geq\frac{a_n}{b_n}\geq 0$;

\item[Case 3.] 
\begin{equation}  \label{case3}
\phi_\sigma(\alpha):=\sum_{i=1}^m a_{i} \delta(\alpha-\alpha_i) +
\sum_{i=m+1}^{n} a_{i} \delta(\alpha-\alpha_i), \quad
\phi_\varepsilon(\alpha):=\sum_{i=m+1}^{n} b_{i} \delta(\alpha-\alpha_i),
\end{equation}
with $0\leq\alpha_1<\ldots<\alpha_m<\alpha_{m+1}<\ldots<\alpha_n<1$, and $%
\frac{a_{m+1}}{b_{m+1}}\geq\frac{a_{m+2}}{b_{m+2}}\geq\ldots\geq\frac{a_n}{%
b_n}\geq 0$;

\item[Case 4.] 
\begin{equation}  \label{case4}
\phi_\sigma(\alpha):=\sum_{i=1}^{n} a_{i} \delta(\alpha-\alpha_i), \quad
\phi_\varepsilon(\alpha):=\sum_{j=1}^{m} b_{j} \delta(\alpha-\beta_j),
\end{equation}
with $\alpha_i\not=\beta_j$, for all $i\not= j$, and $0\leq\alpha_1<\ldots<%
\alpha_n<\beta_1<\ldots<\beta_m< 1$.
\end{itemize}

In all four cases all coefficients $a_i$ and $b_i$ are supposed to be
nonnegative.
\end{itemize}

Following results show that the assumptions \textbf{(A1)}-\textbf{(A6)} are
always satisfied for any of four admissible linear fractional models of the
wave equation described above.

\begin{thm}
\label{th:aA4} Let $u_0, v_0\in \mathcal{S} ^{\prime }(\mathbb{R})$. Let the
constitutive distributions $\phi_\sigma$ and $\phi_\varepsilon$ in the
stress-strain relation \eqref{system-2} be determined by any of the cases %
\eqref{case1}, \eqref{case2}, \eqref{case3} or \eqref{case4}. Then there
exists a unique solution $u\in\mathcal{S} ^{\prime }(\mathbb{R}\times 
\mathbb{R}_{+})$ to the generalized Cauchy problem \eqref{dowe-gCp} given by %
\eqref{fund sol u}.
\end{thm}

In order to prove the theorem, we need the following lemma.

\begin{lem}
\label{lem:polyn1} Let $\tau_i>0$, $i=0,1,\ldots,n$, and $0\leq
\alpha_0<\alpha_1<\ldots<\alpha_n<1$. Then the polynomial function 
\begin{equation*}
f(s) = \sum_{i=1}^{n} \tau_i\, s^{\alpha_i}
\end{equation*}
has no zeros in $\mathbb{C}\backslash \{0\}$.
\end{lem}

\noindent \textbf{Proof. } First notice that $f(\bar{s})=\overline{f(s)}$,
where the bar denotes the complex conjugation, so it is enough to consider $%
s $ only in the upper complex half-plane.

Thus, write $s=re^{i\varphi }$, $r>0$, $\varphi \in (0,\pi )$. Then 
\begin{equation*}
\mathop{\rm Im}\nolimits f=\sum_{i=1}^{n}\tau _{i}\,r^{\alpha _{i}}\sin
(\alpha _{i}\varphi )>0,
\end{equation*}%
since $\sin (\alpha _{i}\varphi )>0$, for $\alpha _{i}\varphi \in (0,\pi )$, 
$i=0,1,\ldots ,n$, and hence $f$ does not vanish. Obviously, the same holds
also when $s$ lies on positive or negative part of the real line, which
proves the claim. \nolinebreak {\hspace*{\fill}$\Box $ \vspace*{0.25cm}}

\noindent \textbf{Proof of Theorem \ref{th:aA4}. } We shall show that the
system of equations \eqref{system-1} - \eqref{system-3} satisfies
assumptions \textbf{(A1)}-\textbf{(A6)}.

In order to prove assumptions \textbf{(A1)} and \textbf{(A3)}, it is enough
to show that the functions $\Phi_\sigma$ and $\Phi_\varepsilon$ introduced
in \eqref{fiovi}, that are polynomials in the present situation, does not
vanish for $\mathop{\rm Re}\nolimits s>0$. It will imply \eqref{cond-inv-L},
and thus the existence of the inverse Laplace transforms. But this follows
from the previous Lemma \ref{lem:polyn1}, since $\Phi_\sigma$ and $%
\Phi_\varepsilon$ are complex polynomials of $s$ of degree less than $1$.
The same argument applies for verifying assumption \textbf{(A4)}, since the
singularities of multiform function $\tilde{S}(x,s)=\frac{1}{2}\sqrt{\frac{%
\Phi_\sigma(s)}{\Phi_\varepsilon(s)}}\, e^{-|x|s\sqrt{\frac{\Phi_\sigma(s)}{%
\Phi_\varepsilon(s)}}} $ coincide with the zeros of $\Phi_\sigma$ and $%
\Phi_\varepsilon$.

Consider 
\begin{equation}
\psi (\xi ,s)=s^{2}+\xi ^{2}\frac{\Phi_{\varepsilon }(s)}{\Phi_{\sigma }(s)}
=s^{2}+\xi ^{2}\frac{\sum_{j=1}^{m}b_{j}s^{\beta_{j}}} {%
\sum_{i=1}^{n}a_{i}s^{\alpha_{i}}},\qquad \xi \in \mathbb{R},\,\,s\in 
\mathbb{C},  \label{psi}
\end{equation}
defined in the assumption \textbf{(A2')}, with $\Phi_{\varepsilon }$ and $%
\Phi_{\sigma }$ obtained by \eqref{fiovi} using the most general form of
constitutive distributions $\phi_{\varepsilon }$ and $\phi_{\sigma }$, given
by \eqref{case4}. Introducing $s=re^{i\varphi }$, $\varphi \in (-\pi,\pi)$,
in \eqref{psi} and by separating real and imaginary parts, one obtains in
each of four cases the following (with $\theta =\frac{\xi ^{2}}{%
|\sum_{i=1}^{n} a_{i}s^{\alpha_{i}}|^{2}}$):

\begin{itemize}
\item[Case 1.] $n=m$ and $\beta_{i}=\alpha_{i}$, $i=1,\ldots ,n$ 
\begin{equation*}
\mathop{\rm Im}\nolimits \psi (\xi ,s) = r^{2}\sin (2\varphi) +\theta
\sum_{j=2}^{n} \sum_{i=1}^{j-1} (a_{i}b_{j}-a_{j}b_{i})
r^{\alpha_{i}+\alpha_{j}} \sin((\alpha_{j}-\alpha_{i})\varphi)>0;
\end{equation*}

\item[Case 2.] $n<m$ and $\beta_{q}=\alpha_{q}$, $q=1,\ldots ,n$ 
\begin{eqnarray*}
\mathop{\rm Im}\nolimits \psi (\xi ,s) &\!\!\!=\!\!\!& r^{2} \sin (2\varphi)
+\theta \sum_{j=2}^{n}\sum_{i=1}^{j-1} (a_{i}b_{j}-a_{j}b_{i})
r^{\alpha_{p}+\alpha_{q}} \sin((\alpha_{j}-\alpha_{i})\varphi) \\
&& \,\, + \theta \sum_{i=1}^{n} \sum_{j=n+1}^{m} a_{i}b_{j}
r^{\alpha_{i}+\beta_{j}}\sin ((\beta_{j}-\alpha_{i})\varphi) >0;
\end{eqnarray*}

\item[Case 3.] $n>m$ with $n=m+m^{\prime }$ and $\beta_{j}=\alpha_{m^{\prime
}+j}$, $j=1,\ldots,m$ 
\begin{eqnarray*}
\mathop{\rm Im}\nolimits \psi (\xi ,s) &\!\!\!=\!\!\!& r^{2}\sin (2\varphi)
+\theta \sum_{i=1}^{m^{\prime }} \sum_{j=1}^{m} a_{i}b_{j}
r^{\alpha_{i}+\alpha_{m^{\prime }+j}} \sin((\alpha_{m^{\prime
}+j}-\alpha_{i})\varphi) \\
&& \,\, + \theta \sum_{j=n+2}^{m} \sum_{j^{\prime }=1}^{j-1} (a_{m^{\prime
}+q^{\prime }}b_{j}-a_{m^{\prime }+j}b_{j^{\prime }}) r^{\alpha_{m^{\prime
}+j}+\alpha_{m^{\prime }+j^{\prime }}} \sin ((\alpha_{m^{\prime
}+j}-\alpha_{m^{\prime }+j^{\prime }})\varphi) >0;
\end{eqnarray*}

\item[Case 4.] $n\not= m$ and $\alpha_{i}\not= \beta_{j}$, $i=1,\ldots, n$, $%
j=1,\ldots, m$ 
\begin{equation*}
\mathop{\rm Im}\nolimits \psi (\xi ,s)=r^{2} \sin (2\varphi) +\theta
\sum_{i=1}^{n} \sum_{j=1}^{m} a_{i}b_{j}r^{\alpha_{i}+\beta_{j}}
\sin((\beta_{j}-\alpha_{i})\varphi)>0.
\end{equation*}
\end{itemize}

Now the thermodynamical constraints imply that the assumption \textbf{(A2')}
is satisfied in all four cases.

Assumption \textbf{(A5)} is satisfied due to the fact that the highest order
of fractional derivatives of strain is always greater or equal to the
highest order of fractional derivatives of stress, and therefore the limit
in \textbf{(A5)} is a nonnegative constant as the modulus of $s$ converges
to infinity. The same fact also implies that the limit in assumption \textbf{%
(A6)} vanishes as $s$ approaches zero. This proves the claim. \nolinebreak{%
\hspace*{\fill}$\Box$ \vspace*{0.25cm}}

Thermodynamically consistent models, grouped in four cases as described
above, display different material properties in creep and stress relaxation.
Namely, for models corresponding to Cases 1 and 3 one obtains the wave speed
by (\ref{ce}). Also, models belonging to Cases 1 and 2 describe solid-like
materials (material creeps to a finite strain, i.e., equilibrium compliance
is finite), while models of Cases 3 and 4 correspond to fluid-like materials
(material creeps to an infinite strain, i.e., equilibrium compliance is
infinite). Moreover, each of the case corresponds to a specific type of
viscoelastic material, see \cite[Table 2.1]{Mai-10}.

\begin{itemize}
\item[Case 1.] The wave speed, calculated according to (\ref{ce}) is 
\begin{equation*}
c=\sqrt{G_{g}} =\sqrt{\lim_{\left\vert s\right\vert \rightarrow \infty }%
\frac{\sum_{i=1}^{n}b_{i}s^{\alpha _{i}}} {\sum_{i=1}^{n}a_{i}s^{\alpha _{i}}%
}}=\sqrt{\frac{b_{n}}{a_{n}}},
\end{equation*}
since $0\leq \alpha _{1}<\ldots <\alpha _{n}<1$. The previous result is
established in \cite{KOZ11}. Also, equilibrium compliance 
\begin{equation*}
J_{e}:=\lim_{t\rightarrow \infty }J\left( t\right) =\lim_{\left\vert
s\right\vert \rightarrow 0}(s\tilde{J}(s)) =\lim_{\left\vert s\right\vert
\rightarrow 0}\frac{\Phi _{\sigma }(s)}{\Phi _{\varepsilon}(s)}
=\lim_{\left\vert s\right\vert \rightarrow 0}\frac{\sum_{i=1}^{n}a_{i}s^{%
\alpha _{i}}} {\sum_{i=1}^{n}b_{i}s^{\alpha _{i}}}=\frac{a_{1}}{b_{1}},
\end{equation*}
is finite implying that the models describe the solid-like material. Having $%
G_{g},J_{e}<\infty$, models correspond to type I viscoelastic materials.

\item[Case 2.] The glass modulus, calculated by (\ref{gege}), is infinite: 
\begin{equation*}
G_{g}=\lim_{\left\vert s\right\vert \rightarrow \infty } \frac{%
\sum_{i=1}^{n}b_{i}s^{\alpha _{i}}+\sum_{i=n+1}^{m}b_{i}s^{\beta _{i}}} {%
\sum_{i=1}^{n}a_{i}s^{\alpha _{i}}}=\infty ,
\end{equation*}
since $0\leq \alpha _{1}<\ldots <\alpha _{n}<\beta _{n+1}<\ldots
<\beta_{m}<1 $, and therefore one has no information on wave speed. Models
again describe the solid-like material, since the equilibrium compliance is
finite 
\begin{equation*}
J_{e}=\lim_{\left\vert s\right\vert \rightarrow 0}\frac{%
\sum_{i=1}^{n}a_{i}s^{\alpha _{i}}} {\sum_{i=1}^{n}b_{i}s^{\alpha_{i}}+%
\sum_{i=n+1}^{m}b_{i}s^{\beta _{i}}}=\frac{a_{1}}{b_{1}}.
\end{equation*}
Models correspond to type III viscoelastic materials, since $G_{g}=\infty$,
and $J_{e}<\infty$.

\item[Case 3.] The wave speed can be calculated by (\ref{ce}) as 
\begin{equation*}
c=\sqrt{G_{g}}= \sqrt{\lim_{\left\vert s\right\vert \rightarrow \infty }%
\frac{\sum_{i=m+1}^{n}b_{i}s^{\alpha _{i}}} {\sum_{i=1}^{m}a_{i}s^{%
\alpha_{i}}+\sum_{i=m+1}^{n}a_{i}s^{\alpha _{i}}}} =\sqrt{\frac{b_{n}}{a_{n}}%
},
\end{equation*}
since $0\leq \alpha _{1}<\ldots <\alpha _{m}<\alpha _{m+1}<\ldots
<\alpha_{n}<1$. The equilibrium compliance 
\begin{equation*}
J_{e}=\lim_{\left\vert s\right\vert \rightarrow 0} \frac{%
\sum_{i=1}^{m}a_{i}s^{\alpha _{i}}+\sum_{i=m+1}^{n}a_{i}s^{\alpha _{i}}} {%
\sum_{i=m+1}^{n}b_{i}s^{\alpha _{i}}}=\infty ,
\end{equation*}
is infinite implying that the models describe the fluid-like material. Due
to $G_{g}<\infty ,$ $J_{e}=\infty$, models correspond to type II
viscoelastic materials.

\item[Case 4.] No information on wave speed can be obtained, since the glass
modulus, calculated by (\ref{gege}), is infinite 
\begin{equation*}
G_{g}=\lim_{\left\vert s\right\vert \rightarrow \infty }\frac{%
\sum_{j=1}^{m}b_{j}s^{\beta _{j}}} {\sum_{i=1}^{n}a_{i}s^{\alpha _{i}}}%
=\infty ,
\end{equation*}
due to $0\leq \alpha _{1}<\ldots <\alpha _{n}<\beta _{1}<\ldots <\beta
_{m}<1 $, and $\alpha _{i}\not=\beta _{j}$, for all $i\not=j$. Models again
describe the fluid-like material, since the equilibrium compliance is
infinite: 
\begin{equation*}
J_{e}=\lim_{\left\vert s\right\vert \rightarrow 0}\frac{%
\sum_{i=1}^{n}a_{i}s^{\alpha _{i}}} {\sum_{j=1}^{m}b_{j}s^{\beta _{j}}}%
=\infty .
\end{equation*}
Having $G_{g}=J_{e}=\infty$, models correspond to type IV viscoelastic
materials.
\end{itemize}


\section{Power type distributed order model}

\label{sec:cm} 

Power type distributed order model of viscoelastic body, given by (\ref{DOCE}%
) in the dimensional form, is a genuine distributed order model. Taking the
time constant on the right hand side of (\ref{DOCE}) to be the time scale,
the (continuous) constitutive functions $\phi _{\sigma }$ and $\phi
_{\varepsilon }$ in the dimensionless form become: 
\begin{equation*}
\phi _{\sigma }(\alpha )=\tau ^{\alpha },\qquad \phi _{\varepsilon }(\alpha
)=1,\qquad 0<\tau <1,\,\,\,0<\alpha <1,
\end{equation*}%
yielding the corresponding constitutive equation in the form 
\begin{equation}
\int_{0}^{1}\tau ^{\alpha }{}_{0}D_{t}^{\alpha }\sigma (x,t)\,d\alpha
=\int_{0}^{1}{}_{0}D_{t}^{\alpha }\varepsilon (x,t)\,d\alpha .  \label{ccce}
\end{equation}%
The generalized Cauchy problem, corresponding to the power type distributed
order wave equation, takes the form 
\begin{equation}
\frac{\partial ^{2}}{\partial t^{2}}u(x,t)=\mathcal{L}^{-1}\bigg(\frac{%
(s-1)\ln (\tau s)}{(\tau s-1)\ln s}\bigg)\ast _{t}\frac{\partial ^{2}}{%
\partial x^{2}}u(x,t)+u_{0}(x)\delta ^{\prime }(t)+v_{0}(x)\delta (t).
\label{ccCp}
\end{equation}

We shall show in the sequel that power type distributed fractional model
satisfies assumptions \textbf{(A1)}, \textbf{(A2)} and \textbf{(A3)}, which
in turn implies existence of a solution. We shall also show that the
solution to \eqref{ccCp} can be explicitly calculated, since assumptions 
\textbf{(A4)}, \textbf{(A5)} and \textbf{(A6)} are satisfied as well.

\begin{thm}
\label{th:ccA123} Suppose $\phi_\sigma (\alpha) = \tau^\alpha$, $%
\phi_\varepsilon(\alpha) = 1$, with $0<\tau<1$, and $u_0, v_0\in \mathcal{S}
^{\prime }(\mathbb{R})$. Then

\begin{itemize}
\item[(i)] System \eqref{system-1}, \eqref{ccce}, \eqref{system-3}, with the
initial conditions \eqref{ic}, and equation \eqref{ccCp}, are equivalent.

\item[(ii)] There exists a unique solution $u\in\mathcal{S} ^{\prime }(%
\mathbb{R}\times\mathbb{R}_+)$ to \eqref{ccCp}, supported in the cone $%
|x|<ct $, and given by \eqref{fund sol u}, with $c=\frac{1}{\sqrt{\tau}}$
being the wave propagation speed. Outside the cone, i.e., for $|x|>ct$, $u=0$%
.
\end{itemize}
\end{thm}

\noindent\textbf{Proof. } (i) According to Theorem \ref{th:eq-s-we}, the
first part of the claim will be proved if we show the assumption \textbf{(A1)%
}. Since $\phi_\sigma$ and $\phi_\varepsilon$ are continuous functions on
the compact interval $[0,1]$, they achieve maximal and minimal values, which
are both strictly positive. Therefore we have 
\begin{equation*}
\bigg| \frac{\Phi_\varepsilon(s)}{\Phi_\sigma(s)} \bigg| = \frac{\Big|%
\mathop{\rm Re}\nolimits \int_0^1 \phi_\varepsilon(\alpha)s^{\alpha}
\,d\alpha + i\mathop{\rm Im}\nolimits \int_0^1
\phi_\varepsilon(\alpha)s^{\alpha} \,d\alpha\Big|}{\Big|\mathop{\rm Re}%
\nolimits \int_0^1 \phi_\sigma(\alpha)s^{\alpha} \,d\alpha +i\mathop{\rm Im}%
\nolimits \int_0^1 \phi_\sigma(\alpha)s^{\alpha} \,d\alpha\Big|} \leq \frac{%
\Big|\max_{\alpha\in[0,1]}\phi_\sigma(\alpha)\Big|}{\Big| \min_{\alpha\in[0,1%
]}\phi_\varepsilon(\alpha)\Big|}=\tau,
\end{equation*}
which is the estimate that guarantee existence of the inverse Laplace
transform of $\Phi_\varepsilon/\Phi_\sigma$. This proves \textbf{(A1)}.

(ii) Next, by showing \textbf{(A2)} and \textbf{(A3)}, due to Theorem \ref%
{th:glavna}, we obtain existence of a unique solution to \eqref{ccCp}, while
by verifying the rest of the assumptions \textbf{(A4)}, \textbf{(A5)} and 
\textbf{(A6)}, according to Theorem \ref{th:racunaljka}, we have that the
unique solution is given by \eqref{fund sol}. Thus, in order to prove the
second claim, we need to show that assumptions \textbf{(A2)}-\textbf{(A6)}
are satisfied.

Assumption \textbf{(A3)} is proved in analogous way as \textbf{(A1)}, by
replacing the roles of $\Phi_\varepsilon$ and $\Phi_\sigma$ in the above
fraction.

In order to prove \textbf{(A2)}, we apply the Laplace transform with respect
to $t$ to \eqref{ccCp}, to obtain \eqref{lt-fzwe-distr} with $%
\Phi_\sigma(s)/\Phi_\varepsilon(s) = \frac{(\tau s-1)\ln s}{(s-1)\ln(\tau s)}
$. Then \textbf{(A2)} reads that $s^2 \frac{(\tau s-1)\ln s}{(s-1)\ln(\tau s)%
}$ can not take negative real values (including zero), or equivalently 
\textbf{(A2')} states that $\psi(\xi,s)=s^2+\xi^2 \frac{(\tau s-1)\ln s}{%
(s-1)\ln(\tau s)} \not= 0$, which in turn guarantee solvability of the wave
equation. In fact, we shall prove \textbf{(A2')}.

It was shown in \cite{a} that the (principal branch of) function $%
F(s)=s^2+\omega^2+\alpha\beta\Phi(s)$ has exactly two zeros which are
simple, conjugate and placed in the open left half-plane, provided that $%
\alpha\beta>0$, $\omega^2\geq 0$. Thus, by choosing $\omega=0$ and $%
\alpha\beta=\xi^2$, we conclude that the same holds for $\psi$, i.e., $%
\psi(\xi,s)\not=0$, for $\xi\in\mathbb{R}$ and $\mathop{\rm Re}\nolimits s>0 
$. By this, the uniqueness of the solution to \eqref{ccCp} is proved.

It remains to show that the unique solution to \eqref{ccCp} is given by %
\eqref{fund sol}. Here, the multiform function $\tilde{K}$ takes the form 
\begin{equation*}
\tilde{K}(x,s)=\frac{1}{2}\sqrt{\frac{(\tau s-1)\ln s}{(s-1)\ln (\tau s)}}%
\,e^{-|x|s\sqrt{\frac{(\tau s-1)\ln s}{(s-1)\ln (\tau s)}}}.
\end{equation*}%
It has no singularities in $\mathbb{C}$, since $\lim_{s\rightarrow 1}\tilde{K%
}(x,s)$ and $\lim_{s\rightarrow 1/\tau }\tilde{K}(x,s)$ exist in $\mathbb{C}$%
. Thus, assumption \textbf{(A4)} is satisfied.

Assumptions \textbf{(A5)} and \textbf{(A6)} can also be verified, since 
\begin{equation*}
\lim_{R\to \infty} \bigg| \sqrt{\frac{\Phi_\sigma (Re^{i\varphi})}{%
\Phi_\varepsilon (Re^{i\varphi})}} \bigg| = \lim_{R\to \infty} \bigg| \sqrt{%
\frac{(\tau Re^{i\varphi}-1)( \ln R + i\varphi)}{(Re^{i\varphi}-1)(\ln \tau
+ \ln R + i\varphi)}} \bigg| = \sqrt{\tau},
\end{equation*}
and 
\begin{equation*}
\lim_{\eta\to 0} \bigg|\eta \sqrt{\frac{\Phi_\sigma (\eta e^{i\varphi})}{%
\Phi_\varepsilon (\eta e^{i\varphi})}} \bigg| = \lim_{\eta\to 0} \bigg| \eta 
\sqrt{\frac{(\tau\eta e^{i\varphi}-1)(\ln\eta + i\varphi)}{(\eta
e^{i\varphi}-1)(\ln \tau + \ln\eta + i\varphi)}} \bigg| =0.
\end{equation*}
Therefore, power type distributed order wave equation has finite wave
propagation speed which equals to $\frac{1}{\sqrt{\tau}}$. This proves the
theorem. \nolinebreak{\hspace*{\fill}$\Box$ \vspace*{0.25cm}}


\section{Numerical experiments}

\label{sec:nums} 

We aim to investigate the propagation of initial Dirac-delta displacement,
for different cases of constitutive equation of viscoelastic body %
\eqref{system-2}. The spatial profiles of solution $u$, given by 
\eqref{fund
sol u}, assuming initial conditions \eqref{ic} as $u_{0}=\delta $ and $%
v_{0}=0$, will be depicted at given time-instances. Actually, the response
to Dirac distribution is the fundamental solution $K$, given by 
\eqref{fund
sol}. In the case of linear fractional models, we shall treat a
representative of each Case 1 - 4, with constitutive distributions taking
the forms \eqref{case1}-\eqref{case4}, and in the case of the distributed
order model, we treat the power type constitutive functions yielding the
model \eqref{ccce}. Note that all models will be given in dimensionless form.

Fractional Zener model 
\begin{equation*}
\sigma (x,t) + \tau\, {}_{0}D_{t}^{\alpha}\sigma (x,t) = \varepsilon (x,t) +
{}_{0}D_{t}^{\alpha}\varepsilon (x,t), \qquad 0<\alpha<1, 0<\tau<1,
\end{equation*}
is a representative of Case 1 (models having the same number and orders of
fractional derivatives of stress and strain in the constitutive equation).
It was studied in details in \cite{KOZ10}, where existence and uniqueness of
the fundamental solution were shown. The form of model belonging to Case 1
with constitutive distribution assumed as \eqref{case1}, thus generalizing
the fractional Zener model, was investigated in \cite{KOZ11}.

Modified Zener, modified Maxwell, and Maxwell fractional models of
viscoelastic body, respectively given by 
\begin{eqnarray}
&& (1 + a\, {}_{0}D_{t}^{\alpha})\sigma (x,t) = (1 + b\,
{}_{0}D_{t}^{\alpha} + {}_{0}D_{t}^{\beta})\varepsilon (x,t), \quad
0<\alpha<\beta<1, 0<a<b,  \label{mfZce} \\
&& (1 + a_0\, {}_{0}D_{t}^{\alpha_0} + a_1\, {}_{0}D_{t}^{\alpha_1})\sigma
(x,t) = {}_{0}D_{t}^{\alpha_1}\varepsilon (x,t), \quad
0<\alpha_0<\alpha_1<1, a_0,a_1>0,  \label{mMm} \\
&& (1 + a\, {}_{0}D_{t}^{\alpha})\sigma (x,t) =
{}_{0}D_{t}^{\beta}\varepsilon (x,t), \quad 0<\alpha<\beta<1, a>0,
\label{Mm}
\end{eqnarray}
are representatives of Case 2 (all orders of fractional derivatives of
stress are the same as of strain, with some extra derivatives of strain),
Case 3 (all orders of fractional derivatives of strain are the same as of
stress, with some extra derivatives of stress), and Case 4 (all orders of
fractional derivatives of stress and strain are different), respectively.

\begin{figure}[p]
\begin{center}
\begin{minipage}{72mm}
   \includegraphics[scale=0.5]{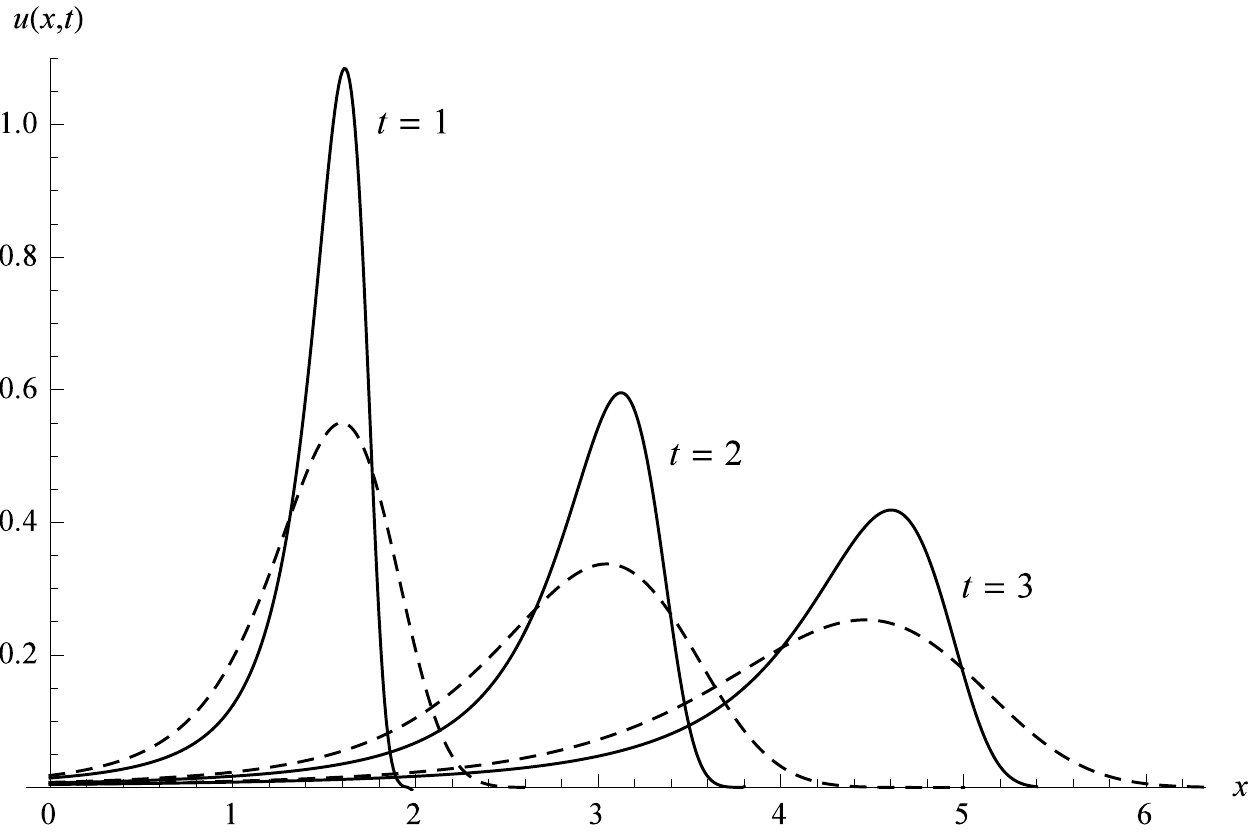}
   \caption*{{\tiny $\alpha=0.25$, $\beta=0.5$ - solid line, $\beta=0.75$ - dashed line}}
  \end{minipage}
\hfil
\begin{minipage}{72mm}
   \includegraphics[scale=0.5]{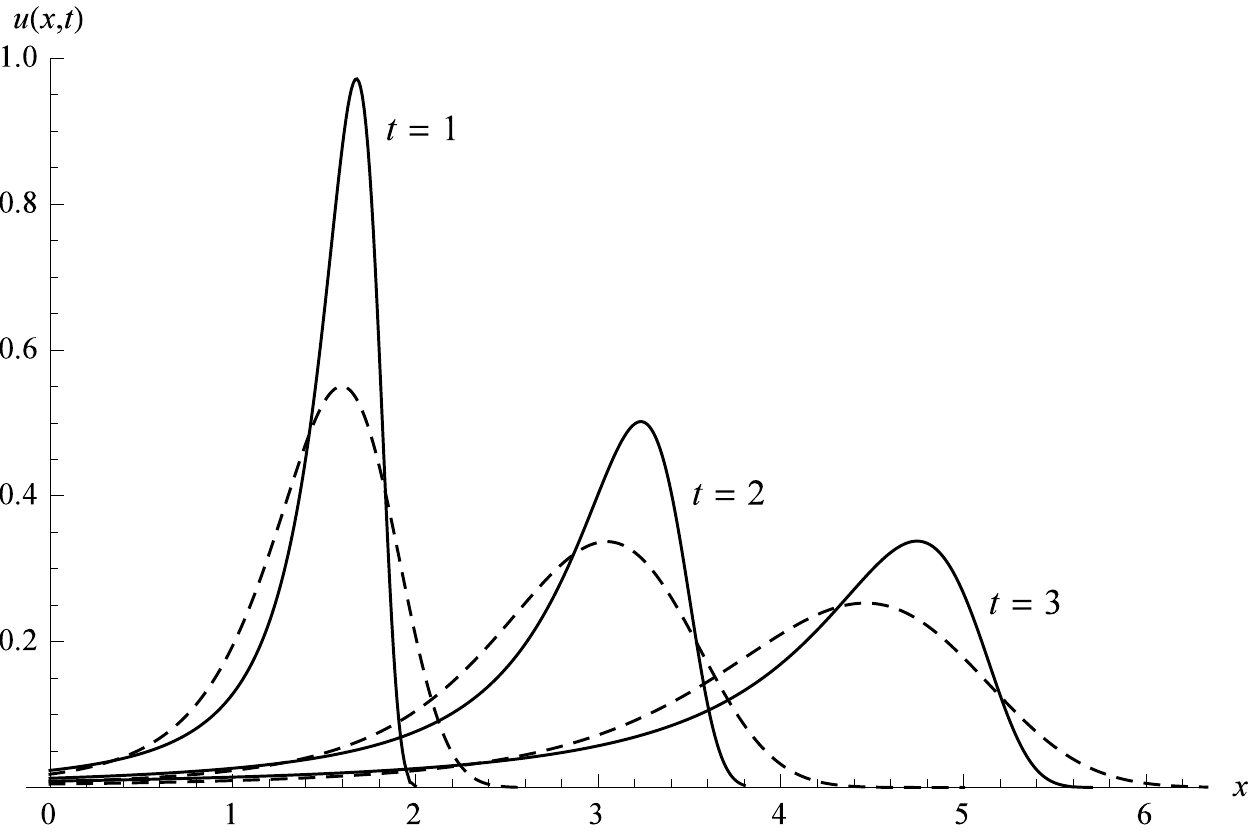}
   \caption*{{\tiny $\beta=0.75$, $\alpha=0.25$ - dashed line, $\alpha=0.5$ - solid line}}
   \end{minipage}
\end{center}
\caption{Spatial profiles of solution at different time-instances - case of
modified Zener model with $a=2$, and $b=4$.}
\label{fig.1}
\end{figure}
\begin{figure}[p]
\begin{center}
\begin{minipage}{72mm}
   \includegraphics[scale=0.5]{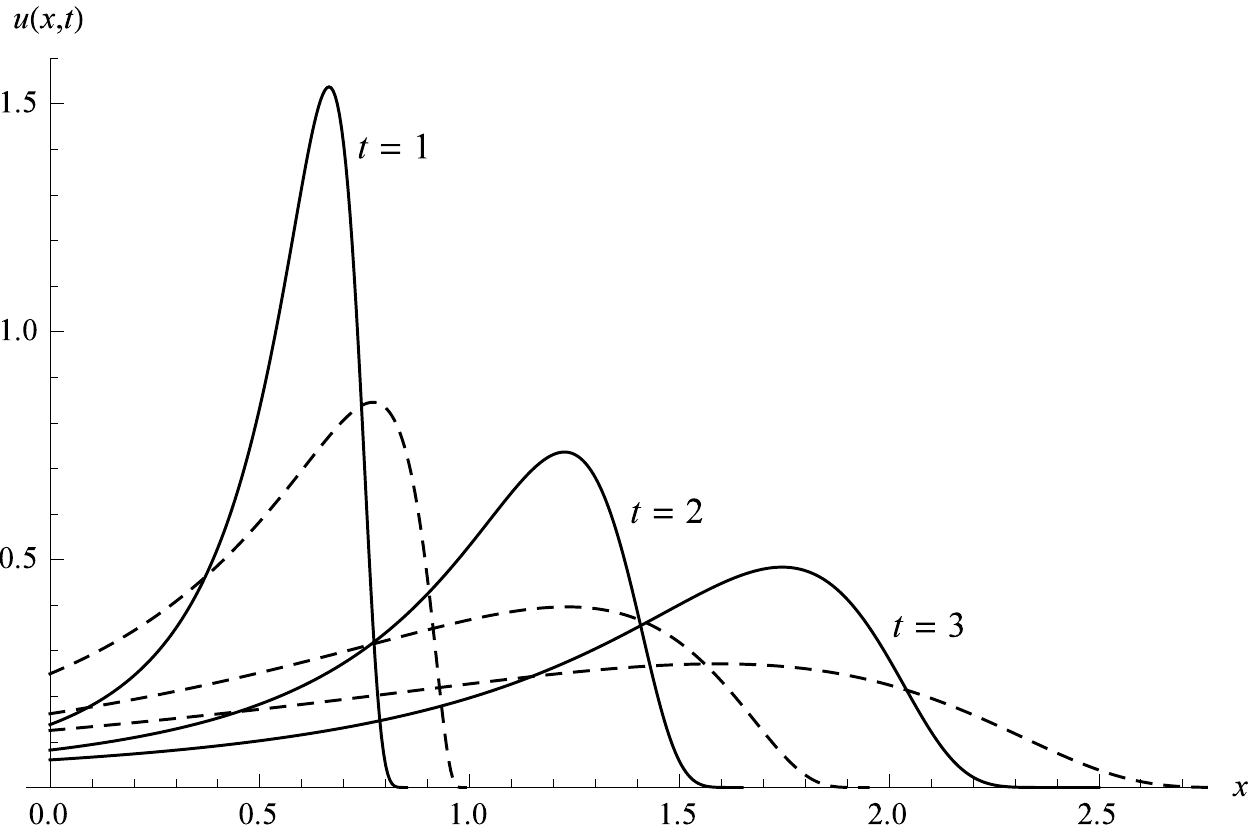}
   \caption*{{\tiny $\alpha_0=0.25$, $\alpha_1=0.5$ - solid line, $\alpha_1=0.75$ - dashed line}}
  \end{minipage}
\hfil
\begin{minipage}{72mm}
    \includegraphics[scale=0.5]{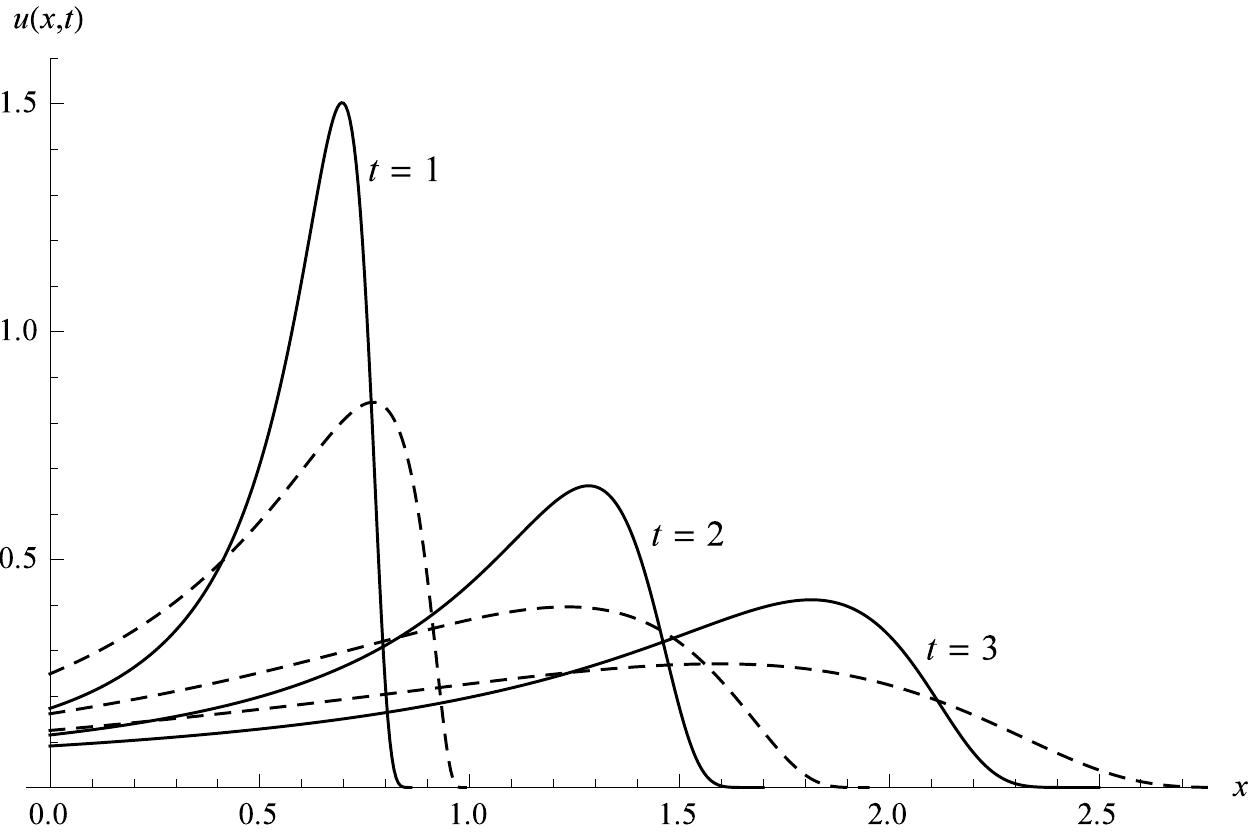}
    \caption*{{\tiny $\alpha_1=0.75$, $\alpha_0=0.25$ - dashed line, $\alpha_0=0.5$ - solid line}}
   \end{minipage}
\end{center}
\caption{Spatial profiles of solution at different time-instances - case of
modified Maxwell model with $a_0=2$, and $a_1=1$.}
\label{fig.2}
\end{figure}
\begin{figure}[p]
\begin{center}
\begin{minipage}{72mm}
   \includegraphics[scale=0.5]{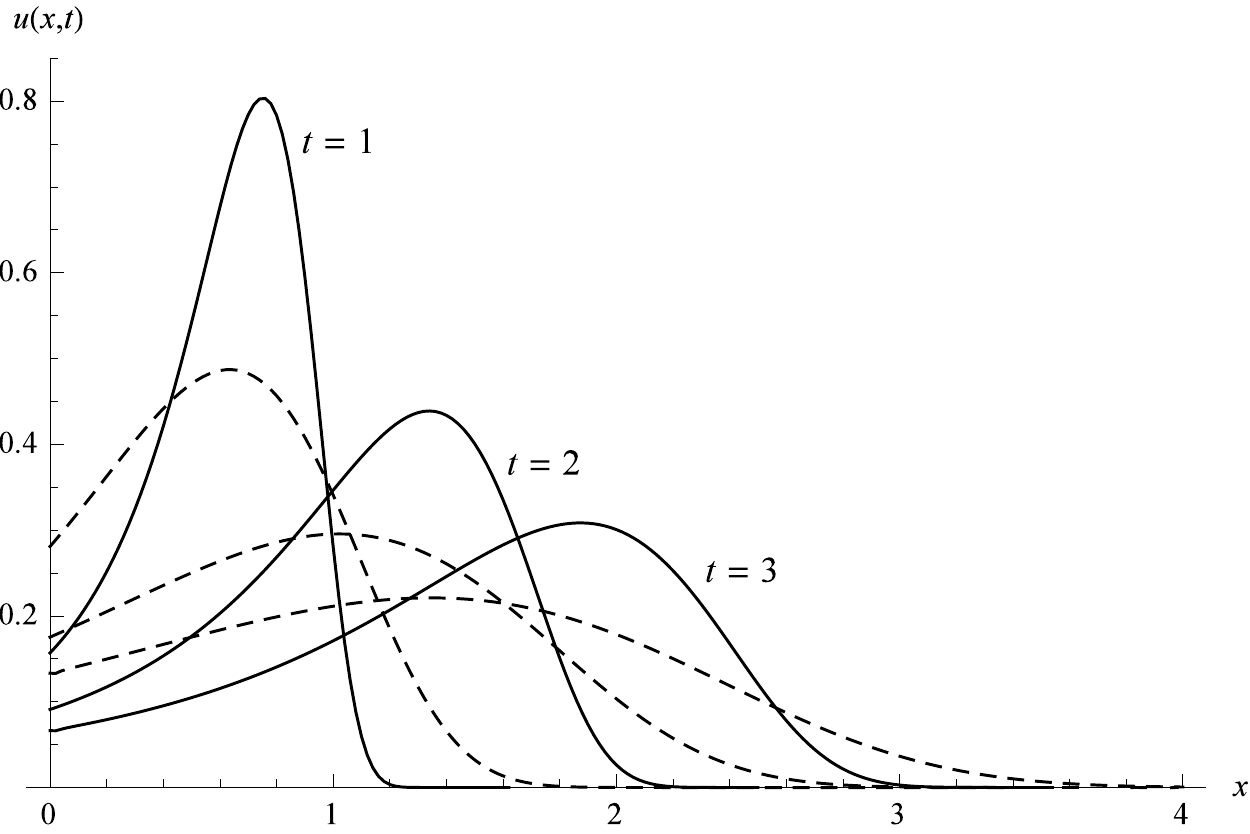}
   \caption*{{\tiny $\alpha=0.25$, $\beta=0.5$ - solid line, $\beta=0.75$ - dashed line}}
  \end{minipage}
\hfil
\begin{minipage}{72mm}
   \includegraphics[scale=0.5]{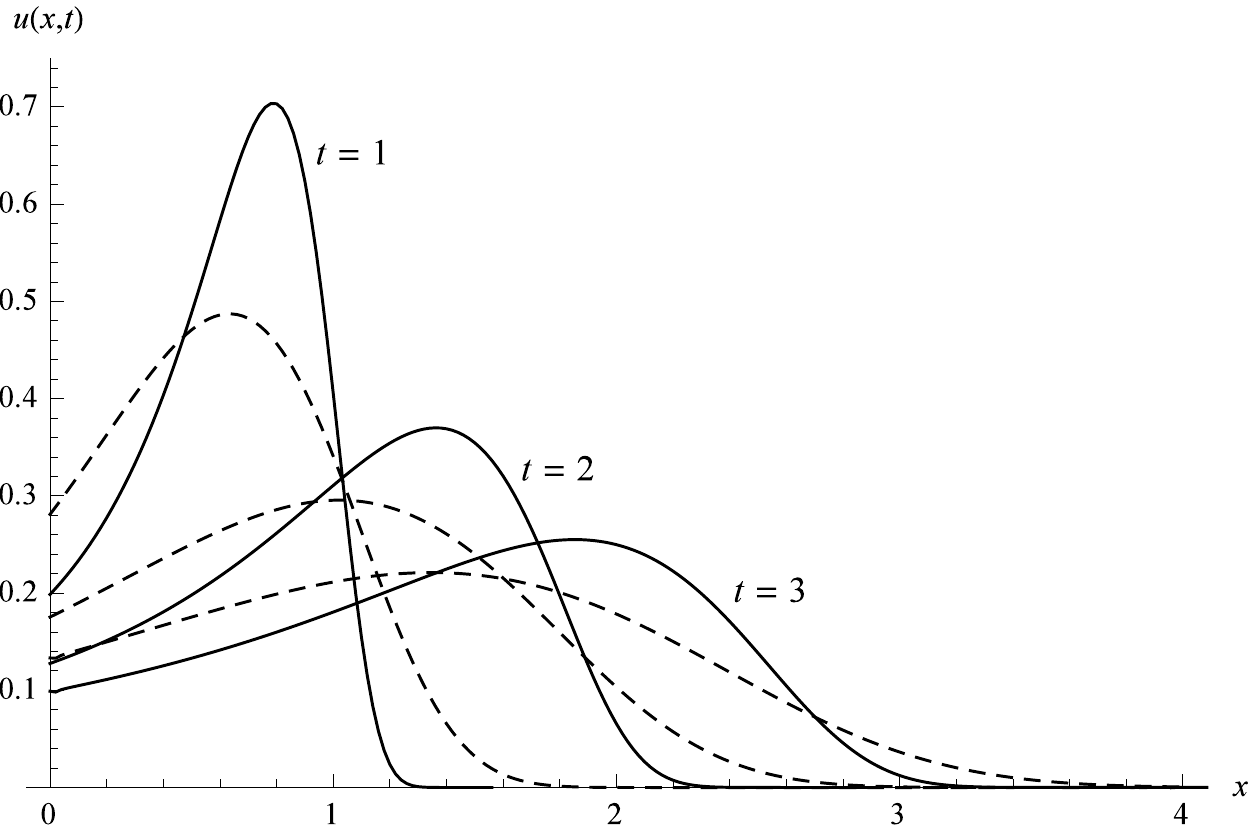}
   \caption*{{\tiny $\beta=0.75$, $\alpha=0.25$ - dashed line, $\alpha=0.5$ - solid line}}
  \end{minipage}
\end{center}
\caption{Spatial profiles of solution at different time-instances - case of
Maxwell model with $a=2$.}
\label{fig.3}
\end{figure}

Figures \ref{fig.1}, \ref{fig.2}, and \ref{fig.3} present the time-evolution
of spatial profile of a material response to initial Dirac-delta
displacement. The material is modeled by constitutive equations \eqref{mfZce}%
, \eqref{mMm}, and \eqref{Mm}, respectively. One notices that the initial
Dirac-delta profile evolves in space during time, as expected in wave
propagation modeling. Due to the energy dissipativity property of the
viscoelastic material, as time passes, the peaks decrease in height and
increase in width. The peak heights and widths are different due to
differences in models and model parameters. For the selected values of
parameters, modified fractional Zener model, since that peaks are quite
narrow, displays properties that are the most similar to the purely elastic
model (when Dirac-delta evolves without any distortions), see Figure \ref%
{fig.1}. The widest peaks are obtained in the case of the fractional Maxwell
model, see Figure \ref{fig.3}. This is expected since the model describes
the fluid-like viscoelastic body.

There two sub-figures containing in each figure. In obtaining graphs
presented in the left-hand-side sub-figure, parameter $\alpha$ ($\alpha_0$)
is fixed and parameter $\beta$ ($\alpha_1$) is varied. One notices that, as $%
\beta$ ($\alpha_1$) is increased peak heights are decreased, while their
width is increased. The cause for such behavior may be the increased energy
dissipativity of the material. The opposite effect is obtained when
increasing the parameter $\alpha$ ($\alpha_0$), while having $\beta$ ($%
\alpha_1$) fixed, as depicted in graphs shown in the right-hand-side
sub-figures. Namely, as $\alpha$ ($\alpha_0$) increases, peak heights
increase, while their width decrease.

\begin{figure}[tbph]
\begin{center}
\begin{minipage}{72mm}
   \includegraphics[scale=0.5]{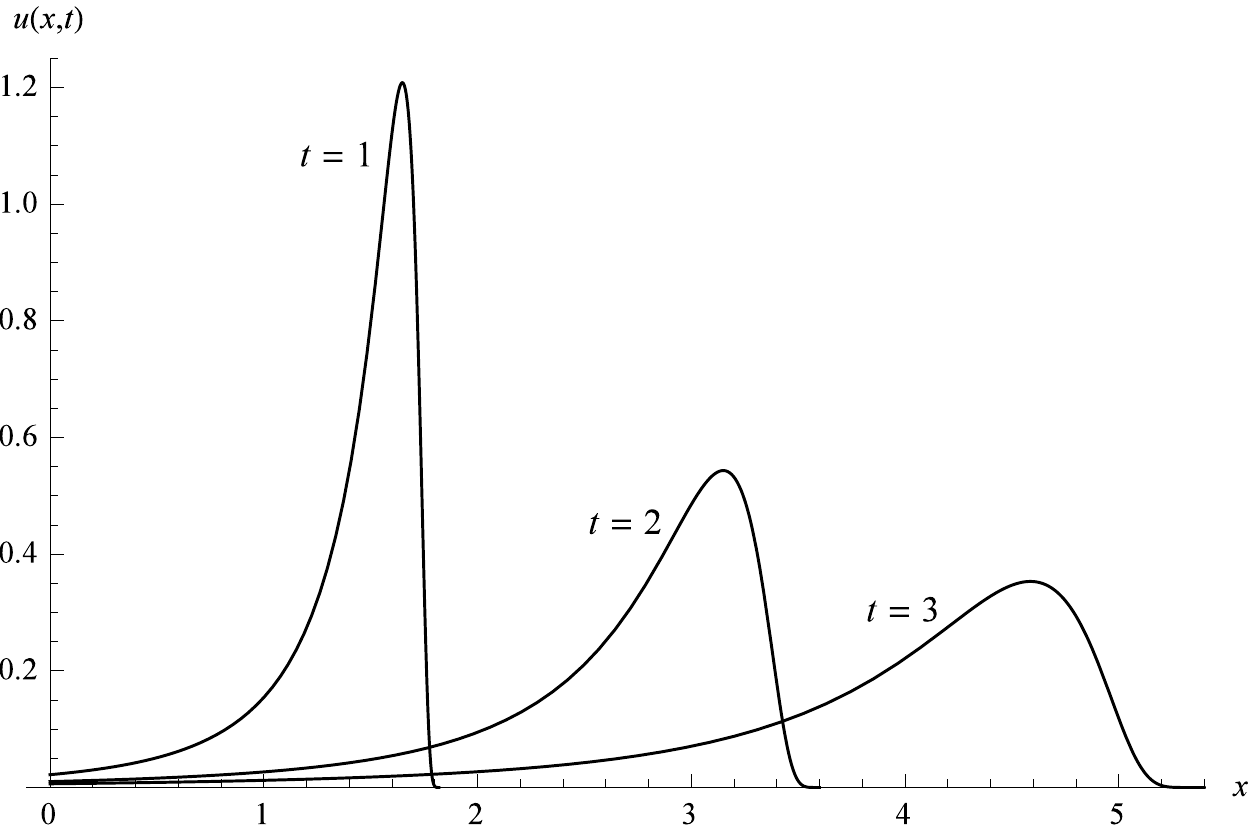}
  \end{minipage}
\hfil
\begin{minipage}{72mm}
   \includegraphics[scale=0.5]{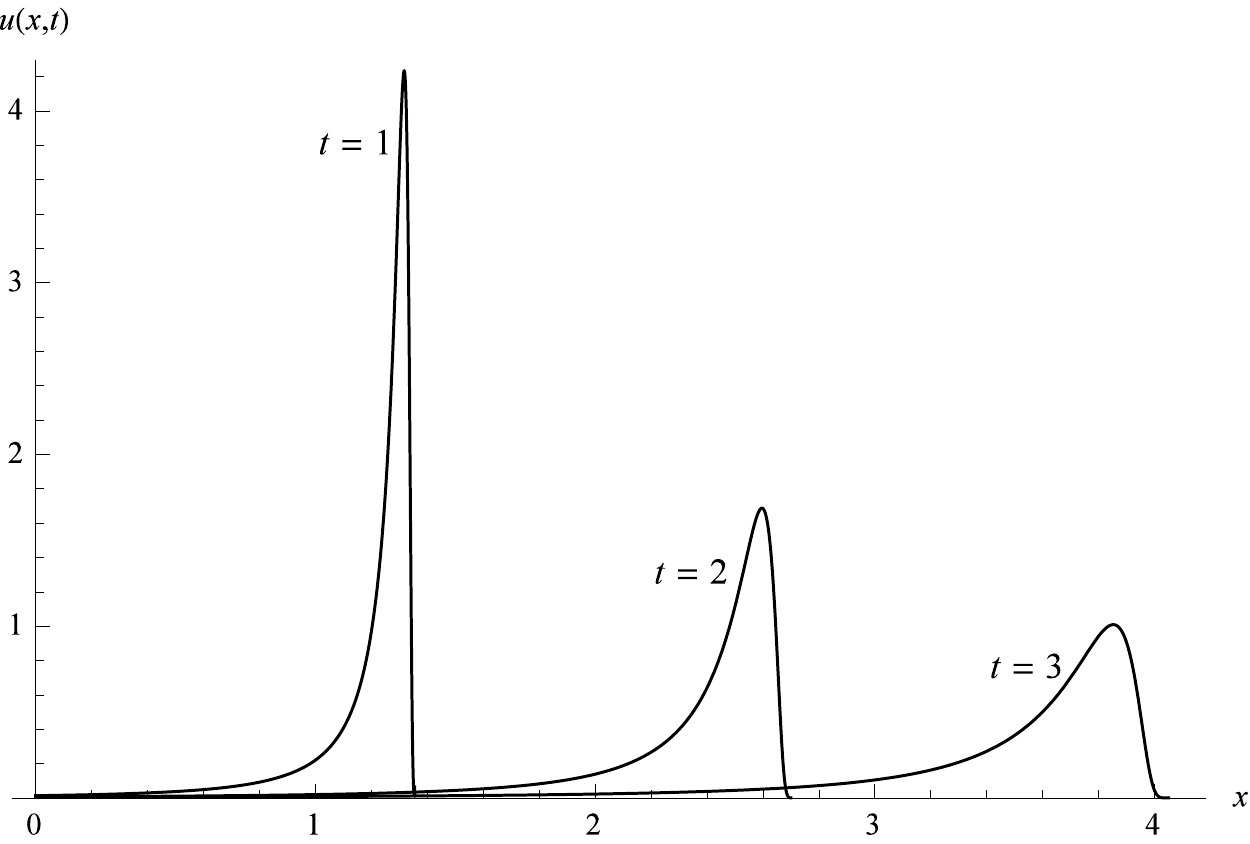}
  \end{minipage}
\vfil
\begin{minipage}{72mm}
   \includegraphics[scale=0.5]{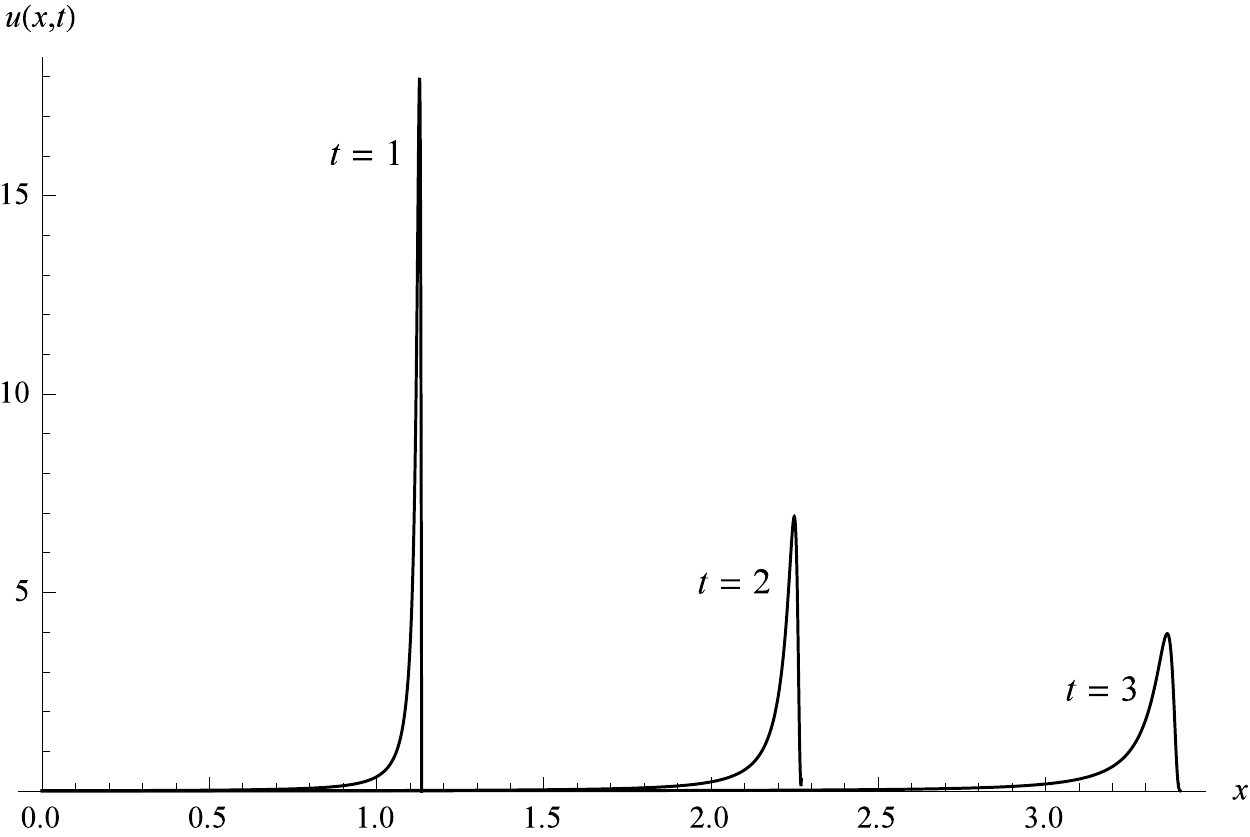}
  \end{minipage}
\end{center}
\caption{Spatial profiles of solution at different time-instances - case of
fractional continuous model with $\protect\tau =0.25$ (upper left), $\protect%
\tau =0.5$ (upper right), and $\protect\tau =0.75$ (lower).}
\label{fig.4}
\end{figure}

Time-evolution of the initial Dirac-delta displacement in the case of
fractional continuous model \eqref{ccce} is presented in Figure \ref{fig.4}
for different values of model parameter $\tau <1$. 
Again, as time passes, peak propagation in space is obtained. Due to the
energy dissipation effect, peak heights decrease, while their width
increases, as the disturbance propagate in space. One notices that for
increased value of parameter $\tau $, at the same time-instances, peak
height is increased and its width is decreased. This behavior is due to the
fact that as the value of parameter $\tau $ increases, the elastic
properties of material become more dominant over the viscous properties. In
the limiting case $\tau =1$, the constitutive equation \eqref{ccce} reduces
to the Hooke law, implying that the response to the initial Dirac-delta
displacement is Dirac-delta as well.


\section*{Conclusion}


Classical wave equation is generalized within the framework of fractional
calculus by considering the distributed order constitutive equation of
viscoelastic deformable body (\ref{const-eq}) instead of the Hooke law,
leaving the basic equations of deformable body: equation of motion (\ref%
{eq-mot}) and strain (\ref{strain}) unchanged. Further, the initial value
Cauchy problem for the distributed order fractional wave equation is
analyzed on the infinite domain within the theory of distributions.

Theorem \ref{th:eq-s-we} provides equivalence of the (dimensionless) system
of equations (\ref{system-1}) - (\ref{system-3}) and equation (\ref{dowe}),
while Theorem \ref{th:glavna} provides conditions that guarantee existence
and uniqueness of a solution to the generalized Cauchy problem (\ref%
{dowe-gCp}). Without specifying the form of constitutive functions or
distribution, three assumptions, guaranteeing the solvability of
constitutive equation with respect to both stress and strain, as well as
fundamental solution's decay to zero at spatial infinity, are required in
order to prove the claims of Theorems \ref{th:eq-s-we} and \ref{th:glavna}.
The fundamental solution is calculated using the Laplace transform inversion
formula in Theorem \ref{th:racunaljka}, that required an additional three
assumptions guaranteeing applicability of the applied method. Two of the
assumptions proved to be connected with the viscoelastic material
properties. Namely, the wave propagation speed is obtained from the support
property of the fundamental solution, and it has proved to be proportional
to the square root of glass modulus. The restriction on growth of the creep
compliance for large times is also the implication of one of the three
assumptions.

Distributed order model (\ref{system-2}) generalizes linear fractional
models of viscoelasticity (\ref{gen-lin}), having the orders of
differentiation in interval $[0,1).$ Theorem \ref{th:aA4} provides existence
and explicit form of the fundamental solution to four classes of fractional
wave equation obtained through thermodynamically admissible models of linear
viscoelasticity, since all six assumptions required by Theorems \ref%
{th:eq-s-we}, \ref{th:glavna}, and \ref{th:racunaljka} are satisfied.
Moreover, it is found that four classes of thermodynamically admissible
linear models correspond to four type of models appearing in viscoelasticity
theory and wave propagation speed is calculated in for the cases when it
defines the conic support of the fundamental solution. Theorem \ref%
{th:ccA123} proves existence and explicit form of the fundamental solution
to the fractional wave equation obtained through the power type distributed
order model (\ref{ccce}), again by showing that all six assumptions are
satisfied. The wave propagation speed is calculated in this case as well.
Numerical examples illustrate propagation of the initial Dirac delta
disturbance in the cases of linear fractional models, as well as in the case
of power type distributed order model.


\section*{Acknowledgement}


This work is supported by Projects $174005$ and $174024$ of the Serbian
Ministry of Education, Science, and Technological Development and Project $%
142-451-2489$ of the Provincial Secretariat for Higher Education and
Scientific Research.



\end{document}